\documentclass[11pt,a4paper]{article}
\pdfoutput=1

\usepackage[nosort]{cite}
\usepackage[bulletsep]{collref}

\usepackage[british]{babel} 

\usepackage{epsfig,rotating}
\usepackage{amsmath,amssymb,amsthm}
\usepackage{amsfonts}
\usepackage{mathrsfs}
\usepackage{bbm}
\usepackage{bm}

\usepackage{color}
\usepackage[textwidth = 430 pt, textheight = 630 pt]{geometry}
\definecolor{MyDarkBlue}{rgb}{0.15,0.25,0.45}

\usepackage[linktocpage=true]{hyperref}
\hypersetup{
colorlinks=true,
citecolor=MyDarkBlue,
linkcolor=MyDarkBlue,
urlcolor=MyDarkBlue,
pdfauthor={Christian S\"amann, Martin Wolf},
pdftitle={Six-Dimensional Superconformal Field Theories\\ from Principal 3-Bundles over Twistor Space},
pdfsubject={hep-th}
breaklinks=true
}

\let\fn\footnote
\renewcommand{\footnote}[1]{\linespread{1.1}\fn{#1}\linespread{1.29}}

\makeatletter\renewcommand{\section}{\@startsection
{section}{1}{\z@}{-3.5ex plus -1ex minus
    -.2ex}{2.3ex plus .2ex}{\bf\mathversion{bold} }}
\makeatletter\renewcommand{\subsection}{\@startsection{subsection}{2}{\z@}{-3.25ex
plus -1ex minus
   -.2ex}{1.5ex plus .2ex}{\bf\mathversion{bold} }}
\makeatletter\renewcommand{\subsubsection}{\@startsection{subsubsection}{3}{-2.45ex}{-3.25ex
plus -1ex minus -.2ex}{1.5ex plus .2ex}{\it }}
\renewcommand{\thesection}{\arabic{section}}
\renewcommand{\thesubsection}{\arabic{section}.\arabic{subsection}}
\renewcommand{\@seccntformat}[1]{\@nameuse{the#1}.~~}

\renewcommand{\theequation}{\thesection.\arabic{equation}}
\makeatletter \@addtoreset{equation}{section}

\renewcommand*\l@section{\@dottedtocline{1}{0em}{2em}}
\renewcommand*\l@subsection{\@dottedtocline{2}{2em}{2.4em}}
\renewcommand*\l@subsubsection{\@dottedtocline{4}{3.8em}{3.7em}}

\renewcommand\tableofcontents{%
    \section*{\large\contentsname
        \@mkboth{%
          \MakeUppercase\contentsname}{\MakeUppercase\contentsname}}%
       {\baselineskip=15pt plus 2pt minus 1pt
    \@starttoc{toc}}%
}

\renewenvironment{thebibliography}[1]
     {\baselineskip=16pt plus 2pt minus 1pt
      \section*{\large\refname
        \@mkboth{\MakeUppercase\refname}{\MakeUppercase\refname}}%
     \list{\@biblabel{\@arabic\c@enumiv}}%
           {\settowidth\labelwidth{\@biblabel{#1}}%
            \leftmargin\labelwidth
            \advance\leftmargin\labelsep
            \@openbib@code
            \usecounter{enumiv}%
            \let\p@enumiv\@empty
            \renewcommand\theenumiv{\@arabic\c@enumiv}}%
      \sloppy
      \clubpenalty4000
      \@clubpenalty \clubpenalty
      \widowpenalty4000%
      \sfcode`\.\@m
 \catcode`\^^M=10%
}

\newcommand{\appendices}{
\section*{Appendix}\label{appendices}\setcounter{subsection}{0}
\addcontentsline{toc}{section}{Appendix}
\setcounter{equation}{0}
\makeatletter
\renewcommand{\theequation}{\Alph{subsection}.\arabic{equation}}
\renewcommand{\thesubsection}{\Alph{subsection}}
\@addtoreset{equation}{subsection}
\makeatother
}

\numberwithin{lemma}{section}

\numberwithin{definition}{section}

\newtheorem{theorem}{Theorem}
\numberwithin{theorem}{section}

\numberwithin{prop}{section}

\numberwithin{cor}{section}

\def\periodb#1{\setbox0=\hbox{$#1$}#1\hskip-\wd0\hbox to\wd0{-}}

\newcommand{\unit}{\mathbbm{1}}   			
\newcommand{\im}{\mathrm{im}}   			

\newcommand{\CF}{\mathcal{F}}

\newcommand{\CH}{\mathcal{H}}

\newcommand{\CN}{\mathcal{N}}

\newcommand{\frg}{\mathfrak{g}}				
\newcommand{\frh}{\mathfrak{h}}				
\newcommand{\frl}{\mathfrak{l}}				

\newcommand{\frS}{\mathfrak{S}}
\newcommand{\frU}{\mathfrak{U}}

\newcommand{\FC}{\mathbbm{C}}     			

\newcommand{\PP}{{\mathbbm{P}}}    			

\newcommand{\dd}{\mathrm{d}}     			
\newcommand{\dpar}{\partial}     			

\newcommand{\eand}{{~~~\mbox{and}~~~}}     		
\newcommand{\ewith}{{~~~\mbox{with}~~~}}

\newcommand{\eforall}{{~~~\mbox{for all}~~~}}

\newcommand{\der}[1]{\frac{\dpar}{\dpar #1}}   		

\newcommand{\au}{\mathfrak{u}}

\newcommand{\sU}{\mathsf{U}}     			

\newcommand{\sAut}{\mathsf{Aut}}

\newcommand{\sSO}{\mathsf{SO}}

\newcommand{\sG}{\mathsf{G}}
\newcommand{\sH}{\mathsf{H}}
\newcommand{\sL}{\mathsf{L}}
\newcommand{\sN}{\mathsf{N}}
\newcommand{\sSp}{\mathsf{Sp}}

\newcommand{\acton}{\vartriangleright}     			
\newcommand{\remark}[1]{}     				
\def\tyng(#1){\hbox{\tiny$\yng(#1)$}}			
\def\tyoung(#1){\hbox{\tiny$\young(#1)$}}			

\newcommand{\sft}{\mathsf{t}}
\newcommand{\sB}{\mathsf{B}}

\begin{document}
\begin{titlepage}

\setcounter{page}{0}
\renewcommand{\thefootnote}{\fnsymbol{footnote}}

\begin{flushright}
 EMPG--13--07\\ DMUS--MP--13/12
\end{flushright}

\begin{center}

{\LARGE\textbf{\mathversion{bold}Six-Dimensional Superconformal Field Theories\\ from Principal 3-Bundles over Twistor Space}\par}

\vspace{1cm}

{\large
Christian S\"amann$^{a}$ and Martin Wolf$^{\,b}$
\footnote{{\it E-mail addresses:\/}
\href{mailto:c.saemann@hw.ac.uk}{\ttfamily c.saemann@hw.ac.uk}, \href{mailto:m.wolf@surrey.ac.uk}{\ttfamily m.wolf@surrey.ac.uk}
}}

\vspace{1cm}

{\it
$^a$ Maxwell Institute for Mathematical Sciences\\
Department of Mathematics,
Heriot--Watt University\\
Edinburgh EH14 4AS, United Kingdom\\[.5cm]

$^b$
Department of Mathematics,
University of Surrey\\
Guildford GU2 7XH, United Kingdom\\[.5cm]

}

\vspace{1cm}

{\bf Abstract}
\end{center}
\vspace{-.3cm}

\begin{quote}

We construct manifestly superconformal field theories in six dimensions which contain a non-Abelian tensor multiplet. In particular, we show how principal 3-bundles over a suitable twistor space encode solutions to these self-dual tensor field theories via a Penrose--Ward transform. The resulting higher or categorified gauge theories significantly generalise those obtained previously from principal 2-bundles in that the so-called Peiffer identity is relaxed in a systematic fashion. This transform also exposes various unexplored structures of higher gauge theories modelled on principal 3-bundles such as the relevant gauge transformations. We thus arrive at the non-Abelian differential cohomology that describes principal 3-bundles with connective structure.

\vfill
\noindent 28th April 2014

\end{quote}

\setcounter{footnote}{0}\renewcommand{\thefootnote}{\arabic{thefootnote}}

\end{titlepage}

\tableofcontents

\bigskip
\bigskip
\hrule
\bigskip
\bigskip

\section{Introduction}

Following the impressive success of M2-brane models in the last few years, there is more and more interest in six-dimensional superconformal field theories that might yield candidate theories for similar M5-brane models. These theories should exhibit $\CN=(2,0)$ supersymmetry and have a tensor multiplet in their field content. The biggest issue in the construction of such theories is to render the tensor fields non-Abelian in a meaningful way. While there are a few ad-hoc prescriptions of how to do this, the geometrically most appealing solution to this problem at the moment seems to be higher gauge theory, see e.g.~Baez \& Huerta \cite{Baez:2010ya} and references therein. 

Higher gauge theory describes consistently the parallel transport of extended objects, just as ordinary gauge theory describes the parallel transport of point-like objects. This description arises from a {\it categorification} of the ingredients of ordinary gauge theory. Roughly speaking, under categorification a mathematical notion is replaced by a category in which the notion's original structure equations hold only up to isomorphisms. When categorifying principal bundles, we replace the gauge or structure groups by so-called Lie 2-groups (certain monoidal tensor categories) and the principal bundles by so-called principal 2-bundles (a non-Abelian generalisation of gerbes).

Once a gauge structure is encoded in a way that allows for a description in terms of \v Cech cochains (as e.g.\ principal bundles or bundle gerbes), we can construct a corresponding field theory using twistor geometric techniques: the twistor space $P^6$ suitable for discussing six-dimensional chiral field theories is well known. It is the space that parametrises totally null 3-planes in six-dimensional space-time \cite{springerlink:10.1007/BF00132253,Hughston:1987aa,Penrose:1986ca}. In addition, a generalisation of the Penrose--Ward transform will map the \v Cech cochains to certain differential forms encoding a categorified connection on six-dimensional space-time that satisfies a set of field equations. 

The Penrose--Ward transform for Abelian gerbes over $P^6$ was discussed in \cite{Saemann:2011nb,Mason:2011nw} (see also \cite{0198535651} for an earlier account and \cite{Mason:2012va} for a supersymmetric extension). It yields $\au(1)$-valued self-dual 3-forms in six dimensions. Besides that, in \cite{Saemann:2011nb,Mason:2011nw,Mason:2012va} also twistor space actions have been formulated that represent the twistor analogue of the space-time actions of Pasti, Sorokin \& Tonin \cite{Pasti:1995ii,Pasti:1995tn,Pasti:1996vs,Pasti:1997gx}. 

More recently, we have presented the extension to the non-Abelian case \cite{Saemann:2012uq}: certain non-Abelian gerbes (or principal 2-bundles) over $P^6$ are mapped under a Penrose--Ward transform to the connective structure of a non-Abelian gerbe on space-time that comes with a self-dual 3-form curvature. Since the twistor space $P^6$ can be straightforwardly extended to the supersymmetric case, a Penrose--Ward transform can also be used to identify non-Abelian $\CN=(2,0)$ and  $\CN=(1,0)$ superconformal field equations containing a  tensor multiplet and to describe solutions to these \cite{Saemann:2012uq}.

The principal 2-bundles that are available in the mathematical literature and that have been developed to the extent necessary for applying our Penrose--Ward transform are relatively restricted. Instead of having a fully categorified Lie group (a so-called {\em weak Lie 2-group}) as a gauge group, they only come with what is known as a {\em strict Lie 2-group}. The latter can be regarded as a Lie crossed module, that is, a 2-term complex $\sH\rightarrow \sG$ of ordinary Lie groups $\sG$ and $\sH$. The 3-form curvature of a principal 2-bundle has to satisfy a condition for the underlying parallel transport along surfaces to be well-defined. For a principal 2-bundles with strict Lie 2-structure group, this implies that the 3-form curvature takes values in the centre of the Lie algebra of $\sH$.

This restriction may be regarded as one of the drawbacks of this type of non-Abelian gerbes, and from a topological perspective, these principal 2-bundles thus appear less interesting. We would like to stress, however, that we expect principal 2-bundles still to be relevant from a physical perspective. For instance, the field equations obtained in \cite{Saemann:2012uq} represent an interacting, non-Abelian set of tensor field equations. Moreover,  in \cite{Palmer:2012ya} the 3-Lie algebra valued tensor field equations of Lambert \& Papageorgakis \cite{Lambert:2010wm} have been recast in a higher gauge theory form based on principal 2-bundles. These equations have been studied in detail (see e.g.\ \cite{Richmond:2012de} and references therein), and clearly contain non-trivial dynamics. In particular, after a dimensional reduction, one recovers five-dimensional maximally supersymmetric Yang--Mills theory.

We can avoid the aforementioned drawback in essentially  two ways that remain manageable with the tools available in the literature: first of all, one can work with infinite-dimensional (Lie) crossed modules such as models of the string 2-group discussed by Baez et~al.~\cite{Baez:2005sn}. This approach has been followed by a number of people, see for example Fiorenza, Sati \& Schreiber \cite{Fiorenza:2012tb}. Note that because our twistor constructions \cite{Saemann:2012uq} do not make use of any explicit properties of finite-dimensional crossed modules, they extend to such infinite-dimensional crossed modules without alteration. In the second approach, one can categorify one step further and employ so-called principal 3-bundles having Lie 3-groups as structure $3$-groups. Here, the 3-form curvature can take values in more general subalgebras which are non-Abelian, in general. For a more detailed discussion of this point in a more general context, see also \cite{Fiorenza:2010mh}. In this paper, we shall develop the latter approach.

Principal 3-bundles come with 1-form, 2-form and 3-form gauge potentials together with 2-form, 3-form and 4-form curvatures satisfying certain compatibility relations. If we draw our motivation for the development of superconformal field theories in six dimensions from M-theory with its 3-form potential, the inclusion of such a 3-from potential in the field theory is rather natural. Further motivation for using principal 3-bundles stems from the recently developed $\CN=(1,0)$ superconformal models \cite{Samtleben:2011fj,Chu:2011fd,Samtleben:2012mi,Akyol:2012cq,Samtleben:2012fb,Bandos:2013jva}, which make use of a 3-form gauge potential.\footnote{Note, however, that it seems clear that these models do not directly fit the picture of higher gauge theory on principal 3-bundles, at least not without extending the gauge Lie 3-algebra to a 3-term $L_\infty$-algebra.} For other recent approaches to defining six-dimensional superconformal theories, see e.g.\ references \cite{Lambert:2010wm,Ho:2011ni,Chu:2012um,Bonetti:2012fn,Chu:2012rk,Bonetti:2012st,Chu:2013hja,Fiorenza:2012tb}. 

Higher gauge theory on principal 3-bundles has been developed to a certain extent, see for example Martins \& Picken \cite{Martins:2009aa}, but various details still remain to be clarified. We therefore have two goals in this paper: firstly, we will derive the equations of motion of six-dimensional superconformal models with manifest  $\CN=(n,0)$ supersymmetry for $n=0,1,2$ and encode their solutions in terms of holomorphic principal 3-bundles on twistor space. In deriving the solutions from this holomorphic data, many properties of higher gauge theory on principal 3-bundles (as e.g.\ the explicit form of finite and infinitesimal gauge transformations) will become evident. Describing these properties together with the differential cohomology underlying principal 3-bundles with connection is then our second goal.

This paper is structured as follows. In Section \ref{sec:GG}, we start by reviewing the categorified groups replacing ordinary gauge groups in our theories. We then present the cocycle description of principal 3-bundles in Section \ref{sec:P3B}, which we will use in a rather involved Penrose--Ward transform in Section \ref{sec:PWTSDF}. In Section \ref{sec:discussion}, we discuss the resulting six-dimensional superconformal field theories in detail. In Section \ref{sec:HGT}, we summarise what we have learnt about principal 3-bundles with connective structure by formulating the underlying non-Abelian differential cohomology. We conclude in Section \ref{sec:Conclusions}.

\section{Gauge structure: Lie 3-groups and Lie 2-crossed modules}\label{sec:GG}

The definition of parallel transport of objects that are not point-like and transform under non-Abelian groups has been a long-standing problem. This problem is closely related to that of defining non-Abelian \v{C}ech (and Deligne) cohomology beyond the cohomology set encoding vector bundles. A way to solving both problems is  by categorifying the usual description of gauge theory in terms of principal bundles as explained, for instance, in Baez \& Huerta \cite{Baez:2010ya}. In particular, we will have to categorify the notion of a structure group of a principal bundle.

As already indicated, the general categorifications of the notion of a Lie group lead to so-called {\em weak Lie $n$-groups}. To our knowledge, the theory of principal $n$-bundles with weak Lie $n$-groups as structure $n$-groups has not been developed, at least not to the degree that our constructions require. We therefore have to restrict ourselves to semistrict Lie 3-groups that are encoded by a Lie 2-crossed module, just as strict Lie 2-groups are encoded by Lie crossed modules. Both Lie crossed modules and Lie 2-crossed modules are very manageable, as they consist of 2-term and 3-term complexes of Lie groups, respectively. 

\subsection{Lie crossed modules and differential crossed modules}

In this section, we would like to review briefly the definitions of Lie crossed modules and their infinitesimal version. For more details, see Baez \& Lauda \cite{Baez:0307200} and references therein.

\paragraph{Lie crossed modules.}
Let $(\sG,\sH)$ be a pair of Lie groups. We call the pair $(\sG,\sH)$ a {\it Lie crossed module} if, in addition, there is a smooth $\sG$-action $\acton$ on $\sH$ by automorphisms\footnote{That is, $g\acton(h_1h_2)=(g\acton h_1)(g\acton h_2)$ for all $g\in\sG$ and  $h_1,h_2\in\sH$ and furthermore, $g_1g_2\acton h=g_1\acton(g_2\acton h)$ for all $g_1,g_2\in\sG$ and $h\in\sH$.} (and another one on $\sG$ by conjugation)  and a Lie group homomorphism $\sft:\sH\to\sG$ such that the following two axioms are satisfied:

\begin{center}
\begin{minipage}{14cm}
\begin{itemize}
 \setlength{\itemsep}{-1mm}
\item[(i)] The Lie group homomorphism  $\sft:\sH\to\sG$ is a $\sG$-homomorphism, that is,
 $\sft(g\acton h)=g\sft(h)g^{-1}$ for all $g\in\sG$ and $h\in\sH$.
\item[(ii)] The $\sG$-action $\acton$ and the Lie group homomorphism  $\sft:\sH\to\sG$  obey the so-called {\it Peiffer identity}, $\sft(h_1)\acton h_2=h_1 h_2 h_1^{-1}$ for all $h_1,h_2\in\sH.$
\end{itemize}
\end{minipage}
\end{center}

\noindent
In the following, we shall write $(\sH\overset{\sft}{\to}\sG,\acton)$ or simply $\sH\rightarrow \sG$ to denote a Lie crossed module. Note that Lie crossed modules are in one-to-one correspondence with so-called {\it strict Lie 2-groups} \cite{Baez:0307200}. 

A simple example of a Lie crossed module is $(\sN\overset{\sft}{\to}\sG,\acton)$, where $\sN$ is a normal Lie subgroup of the Lie group $\sG$, $\sft$ is the inclusion, and $\acton$ is conjugation. Another example appears in the non-Abelian gerbes of Breen \& Messing \cite{Breen:math0106083} for which the Lie crossed module is the {\it automorphism Lie 2-group} $(\sG\overset{\sft}{\to}\sAut(\sG),\acton)$ of a Lie group $\sG$, where $\sft$ is the embedding via conjugation and $\acton$ the identity.

\paragraph{Differential Lie crossed modules.}
The infinitesimal counterpart of a Lie crossed module is a differential Lie crossed module. In particular, if $(\frg,\frh)$ is a pair of Lie algebras together with a $\frg$-action $\acton$ on $\frh$ by derivations\footnote{That is, if we denote by $[\cdot,\cdot]$ the Lie brackets on $\frg$ and $\frh$, respectively, then $X\acton [Y_1,Y_2]=[X\acton Y_1,Y_2]+[Y_1,X\acton Y_2]$ for all $X\in\frg$ and all $Y_{1,2}\in\frh$ and furthermore, $[X_1,X_2]\acton Y=X_1\acton(X_2\acton Y)-X_2\acton(X_1\acton Y)$ for all $X_{1,2}\in\frg$ and $Y\in\frh$.} (and on $\frg$ by the adjoint representation)  and a Lie algebra homomorphism $\sft:\frh\to\frg$, then we call  $(\frh\overset{\sft}{\to}\frg,\acton)$ a {\it differential Lie crossed module} provided the linearisations of the two axioms of a Lie crossed module are satisfied:

\begin{center}
\begin{minipage}{14cm}
\begin{itemize}
 \setlength{\itemsep}{-1mm}
\item[ (i)] The Lie algebra homomorphism  $\sft:\sH\to\sG$ is a $\frg$-homomorphism, $\sft(X\acton Y)=[X,\sft(Y)]$ for all $X\in\frg$ and $Y\in\frh$, where $[\cdot,\cdot]$ denotes the Lie bracket on $\frg$.
\item[(ii)] The $\frg$-action $\acton$ and the Lie algebra homomorphism  $\sft:\frh\to\frg$  obey the {\it Peiffer identity},
 $\sft(Y_1)\acton Y_2=[Y_1, Y_2]$ for all $Y_{1,2}\in\frh$, where $[\cdot,\cdot]$ denotes the Lie bracket on $\frh$.
\end{itemize}
\end{minipage}
\end{center}

\noindent
We shall again use $\frh\rightarrow \frg$ as a shorthand notation to denote a differential Lie crossed module. In categorical language, differential Lie crossed modules are in one-to-one correspondence with {\it strict Lie 2-algebras}. We would like to point out that, as shown in \cite{Palmer:2012ya}, the 3-algebras underlying the recently popular M2-brane models can be naturally described in terms of differential Lie crossed modules.

\subsection{Lie 2-crossed modules and differential 2-crossed modules}

As indicated, Lie crossed modules $(\sH\overset{\sft}{\to}\sG,\acton)$ are used as structure $2$-groups in the theory of principal 2-bundles, in the same way that Lie groups are the structure groups for principal bundles (that is, principal 1-bundles). The connective structure on a principal 2-bundle, encoded in a $\frg$-valued connection 1-form and a $\frh$-valued connection 2-form, is, however, somewhat restricted: its associated 3-form curvature has to take values in the kernel of $\sft$ for a consistent parallel transport. Together with the Peiffer identity, this implies that the 3-form curvatures lies in the centre of $\frh$.

We now wish to remove this undesirable restriction by moving away from Lie crossed modules and turning to a categorification of them, the so-called Lie 2-crossed modules of Conduch\'e \cite{Conduche:1984:155} together with their differential counterparts.\footnote{Another categorification of Lie crossed modules is given by Lie crossed squares and Breen \cite{breen-2006} has constructed higher principal bundles using those as structure groups. These crossed squares can be reduced to 2-crossed modules and from a categorical perspective, the latter are sufficiently general.} As we shall see later, these will be used as structure $3$-groups in the theory of principal 3-bundles. Our main motivation to consider this specific generalisation is that the Peiffer identity can be relaxed in a systematic way by means of the so-called {\it Peiffer lifting}. As a direct consequence, the condition that the 3-form curvature takes values in the centre of some Lie algebra will be relaxed, too. This will eventually enable us to construct superconformal self-dual tensor theories in six dimensions with a general 3-form curvature. 

\paragraph{Lie 2-crossed modules.}
Let $(\sG,\sH,\sL)$ be a triple of Lie groups. A {\it Lie 2-crossed module}~\cite{Conduche:1984:155} is a normal complex\footnote{The mappings $\sft$ are Lie group homomorphisms with $\sft^2(\ell)=\unit$ for all $\ell\in\sL$, the images of the mappings $\sft$ are normal Lie subgroups, and $\im(\sft:\sL\to\sH)$ is normal in $\ker(\sft:\sH\to\sL)$. This is a necessary requirement for defining cohomology groups via coset spaces.} of Lie groups,
\begin{equation}
 \sL\ \xrightarrow{~\sft~}\ \sH\ \xrightarrow{~\sft~}\ \sG~,
\end{equation}
together with smooth $\sG$-actions on $\sH$ and $\sL$ by automorphisms (and on $\sG$ by conjugation), both denoted by $\acton$, and a $\sG$-equivariant smooth mapping from $\sH\times\sH$ to $\sL$  called the {\it Peiffer lifting} and denoted by $\{\cdot,\cdot\}: \sH\times \sH\rightarrow \sL$,\footnote{$\sG$-equivariance of $\{\cdot,\cdot\}$ means that $g\acton\{h_1,h_2\}=\{g\acton h_1,g\acton h_2\}$ for all $g\in\sG$ and $h_1,h_2\in\sH$.} subject to the following six axioms (see e.g.~\cite{Conduche:1984:155,Martins:2009aa}):

\begin{center}
\begin{minipage}{14cm}
\begin{itemize}
 \setlength{\itemsep}{-1mm}
 \item[(i)] The Lie group homomorphisms $\sft$ are $\sG$-homomorphisms, that is, $\sft(g\acton \ell)=g\acton \sft(\ell)$ and $\sft(g\acton h)=g\sft(h)g^{-1}$ for all $g\in\sG$, $h\in\sH$, and $\ell\in\sL$.
 \item[(ii)] $\sft(\{h_1,h_2\})=h_1 h_2 h_1^{-1}(\sft(h_1)\acton h_2^{-1})$ for all $h_1,h_2\in\sH$. We define $\langle h_1,h_2\rangle:=h_1 h_2 h_1^{-1}(\sft(h_1)\acton h_2^{-1})$.
 \item[(iii)] $\{\sft(\ell_1),\sft(\ell_2)\}=\ell_1\ell_2\ell_1^{-1}\ell_2^{-1}$ for all $\ell_1,\ell_2\in\sL$. We define $[\ell_1,\ell_2]:=\ell_1\ell_2\ell_1^{-1}\ell_2^{-1}$ (and likewise for elements of $\sG$ and $\sH$).
\end{itemize}
\end{minipage}
\end{center}
\begin{center}
\begin{minipage}{14cm}
\begin{itemize}
 \setlength{\itemsep}{-1mm}
 \item[(iv)] $\{h_1 h_2,h_3\}=\{h_1,h_2h_3h_2^{-1}\}(\sft(h_1)\acton\{h_2,h_3\})$ for all $h_1,h_2,h_3\in\sH$.
 \item[(v)] $\{h_1,h_2h_3\}=\{h_1,h_2\}\{h_1,h_3\}\{\langle h_1,h_3\rangle^{-1},\sft(h_1)\acton h_2\}$ for all $h_1,h_2,h_3\in\sH$.
 \item[(vi)] $\{\sft(\ell),h\}\{h,\sft(\ell)\}=\ell (\sft(h)\acton \ell^{-1})$ for all $h\in\sH$ and $\ell\in\sL$.
\end{itemize}
\end{minipage}
\end{center}

\noindent
We shall write $(\sL\overset{\sft}{\to}\sH\overset{\sft}{\to}\sG,\acton,\{\cdot,\cdot\})$ or, as shorthand notation, $\sL\rightarrow \sH\rightarrow \sG$ to denote a Lie 2-crossed module.  From a categorical point of view, Lie 2-crossed modules encode semistrict Lie 3-groups called Gray groups, see Kamps \& Porter \cite{Kamps:2002aa}, i.e.\ Gray groupoids with a single object.

Furthermore, there is an additional, natural $\sH$-action on $\sL$, also denoted by $\acton$, that is  induced by the above structure,
\begin{equation}\label{eq:ActHToL}
 h\acton \ell\ :=\ \ell\{\sft(\ell)^{-1},h\}\eforall h\ \in\ \sH\eand \ell\ \in\ \sL~.
 \end{equation}
This action is an $\sH$-action on $\sL$ by automorphisms \cite{Conduche:1984:155,Brown:1989aa}, and we recall the proof in Appendix \ref{sec:Proofs}. We directly conclude that
\begin{equation}
  g\acton(h\acton\ell)\ =\ (g\acton h)\acton (g\acton \ell)\eforall g\ \in\sG~,\quad h\ \in\ \sH~,\eand\ell\ \in\ \sL~.
\end{equation}

We would like to emphasise that Lie crossed modules are special instances of Lie 2-crossed modules, and, as such, the latter form natural generalisations of the former. This can be seen as follows. Firstly, let us consider the situation when $\sL$ is the trivial group $\sL=\{\unit\}$. Then, the data $(\sH\overset{\sft}{\to}\sG,\acton)$ form a Lie crossed module since $\{h_1,h_2\}=\unit$ for all $h_1,h_2\in\sH$ and the above axioms (i)--(vi) straightforwardly reduce to those of a Lie crossed module. Secondly, for any Lie 2-crossed module $(\sL\overset{\sft}{\to}\sH\overset{\sft}{\to}\sG,\acton,\{\cdot,\cdot\})$, the truncation $(\sL\overset{\sft}{\to}\sH,\acton)$ with the induced $\sH$-action  \eqref{eq:ActHToL}  forms a Lie crossed module, since from \eqref{eq:ActHToL} and axioms (ii) and (iii) it immediately follows that
\begin{equation}
\begin{aligned}
\sft(h\acton \ell)\ &=\ \sft(\ell)\sft(\{\sft(\ell^{-1}),h\})\ =\ \sft(\ell)\sft(\ell^{-1})h\sft(\ell)h^{-1}\ =\ h\sft(\ell)h^{-1}~,\\
 \sft(\ell_1)\acton\ell_2\ &=\ \ell_2\{\sft(\ell_2^{-1}),\sft(\ell_1)\}\ =\ \ell_1\ell_2\ell_1^{-1}~.
\end{aligned}
\end{equation}
These are precisely the two axioms of a Lie crossed module. Finally, the data $(\sft(\sL)\backslash\sH\overset{\sft}{\to}\sG,\acton)$ also forms a Lie crossed module.

\paragraph{Differential Lie 2-crossed modules.}
As before, we may study the infinitesimal counterpart of Lie 2-crossed modules. Let $(\frg,\frh,\frl)$ be a triple of Lie algebras. A {\it differential Lie 2-crossed module} is a normal complex of Lie algebras
\begin{equation}
 \frl\ \xrightarrow{~\sft~}\ \frh\ \xrightarrow{~\sft~}\ \frg~,
\end{equation}
equipped with $\frg$-actions on $\frh$ and $\frl$ by derivations (and on $\frg$ by the adjoint representation), again denoted by $\acton$, respectively, and a $\frg$-equivariant bilinear map, called again {\em Peiffer lifting} and denoted by $\{\cdot,\cdot\}: \frh\times \frh\rightarrow \frl$, all of which satisfy the following axioms (here, $[\cdot,\cdot]$ denotes the Lie bracket in the respective Lie algebra):

\begin{center}
\begin{minipage}{14cm}
\begin{itemize}
 \setlength{\itemsep}{-1mm}
 \item[(i)] The Lie algebra homomorphisms $\sft$ are $\frg$-homomorphisms, that is, $\sft(X\acton Z)=X\acton\sft(Z)$ and $\sft(X\acton Y)=[X,\sft(Y)]$ for all $X\in\frg$, $Y\in\frh$, and $Z\in\frl$.
 \item[(ii)] $\sft(\{Y_1,Y_2\})=[Y_1,Y_2]-\sft(Y_1)\acton Y_2=:\langle Y_1,Y_2\rangle$ for all $Y_{1,2}\in\frh$.
 \item[(iii)] $\{\sft(Z_1),\sft(Z_2)\}=[Z_1,Z_2]$ for all $Z_{1,2}\in\frl$.
 \item[(iv)] $\{[Y_1,Y_2],Y_3\}=\sft(Y_1)\acton\{Y_2,Y_3\}+\{Y_1,[Y_2,Y_3]\}-\sft(Y_2)\acton\{Y_1,Y_3\}-\{Y_2,[Y_1,Y_3]\}$ for all $Y_{1,2,3}\in\frh$.
 \item[(v)] $\{Y_1,[Y_2,Y_3]\}=\{\sft(\{Y_1,Y_2\}),Y_3\}-\{\sft(\{Y_1,Y_3\}),Y_2\}$ for all $Y_{1,2,3}\in\frh$.
 \item[(vi)] $\{\sft(Z),Y\}+\{Y,\sft(Z)\}=-\sft(Y)\acton Z$  for all $Y\in\frh$ and $Z\in\frl$.
\end{itemize}
\end{minipage}
\end{center}

\noindent
We shall write $(\frl\overset{\sft}{\to}\frh\overset{\sft}{\to}\frg,\acton,\{\cdot,\cdot\})$ or, more succinctly, $\frl\rightarrow \frh\rightarrow \frg$ to denote a differential Lie 2-crossed module. 

Analogously to Lie 2-crossed modules, there is an induced $\frh$-action $\acton$ on $\frl$ that is defined as 
\begin{equation}\label{eq:InducedActionAlgebra}
Y\acton Z\ :=\ -\{\sft(Z),Y\}\eforall Y\ \in\ \frh\eand Z\ \in\ \frl
\end{equation}
and acts by derivations. This $\frh$-action simply follows from the linearisation of \eqref{eq:ActHToL}. In addition, as is a direct consequence of the finite case, $(\frl\overset{\sft}{\to}\frh,\acton)$ forms a differential Lie crossed module.

\section{Principal 3-bundles}\label{sec:P3B}

The next step in our discussion is the introduction of categorified principal bundles that are modelled on Lie 2-crossed modules. Such bundles are called {\em principal 3-bundles}. Following the conventional nomenclature of ordinary principal (1-)bundles, we shall refer to the Lie 2-crossed modules on which the principal 3-bundles are based as the structure $3$-groups. Notice that principal 1-bundles and 2-bundles as well as Abelian 1-gerbes and 2-gerbes will turn out to be special instances of principal 3-bundles.

\subsection{Cocycle description of principal 1- and 2-bundles}

Firstly, let us briefly recall the formulation of smooth principal 1- and 2-bundles in terms of \v{C}ech cohomology. To this end, let $M$ be a smooth manifold and let $\{U_a\}$ be a covering of $M$ which is chosen to be sufficiently fine. In the following, the intersections of coordinate patches that appear will always be assumed to be non-empty.

\paragraph{Principal bundles.}
As is well-known, smooth principal bundles over $M$ with structure group $\sG$ are described in terms of the non-Abelian \v Cech cohomology $H^1(M,\sG)$.\footnote{In writing $H^1(M,\sG)$, we do not make a notational distinction between the Lie group $\sG$ and the sheaf  $\frS_\sG$ of smooth $\sG$-valued functions. We shall continue to use a similar notation when dealing with principal 2-bundles and 3-bundles, respectively.} Representatives of elements of $H^1(M,\sG)$ are 1-cocyles and are called {\em transition functions}. Specifically, a 1-cocycle is a collection $\{g_{ab}\}$ of smooth maps $g_{ab}:U_a\cap U_b\to \sG$ on non-empty intersections  $U_a\cap U_b$ which obey the cocycle condition
\begin{equation}\label{eq:transition_function_principal_1_bundle}
g_{ab}g_{bc}\ =\ g_{ac}\quad\mbox{on}\quad U_a\cap U_b\cap U_c~.
\end{equation}
Note that this condition implies that $g_{aa}=\unit$. Moreover, two principal bundles with transition functions $\{g_{ab}\}$ and $\{\tilde{g}_{ab}\}$ are considered topologically equivalent (or cohomologous), if there are smooth maps $g_a:U_a\rightarrow \sG$ such that
\begin{equation}
 \tilde{g}_{ab}\ =\ g_a^{-1}g_{ab}g_b~.
\end{equation}
A {\it trivial principal bundle} is a principal bundle that is described by transition functions  that are all cohomologous to one, that is, $g_{ab}\sim\unit$ on all $U_a\cap U_b$.

\paragraph{Principal 2-bundles.}
Similarly, smooth principal 2-bundles with structure $2$-groups $(\sH\overset{\sft}{\to}\sG,\acton)$ can be described in terms of a generalised, non-Abelian \v Cech cohomology denoted by $H^2(M,\sH\rightarrow\sG)$. Representatives of elements of this cohomology set are 2-cocyles and are again called transition functions. Specifically, a 2-cocycle is a pair $(\{g_{ab}\},\{h_{abc}\})$ of collections of smooth maps $g_{ab}:U_a\cap U_b\to \sG$ and $h_{abc}:U_a\cap U_b\cap U_c\to \sH$ which obey the following cocycle conditions  \cite{Breen:math0106083,Aschieri:2003mw} (see also \cite{Wockel:2008aa}): 
\begin{equation}\label{eq:transition_function_principal_2_bundle}
\begin{aligned}
\sft(h_{abc})g_{ab}g_{bc}\ &=\ g_{ac}\quad\mbox{on}\quad U_a\cap U_b\cap U_c~,\\
h_{acd}h_{abc}\ &=\ h_{abd}(g_{ab}\acton h_{bcd})\quad\mbox{on}\quad U_a\cap U_b\cap U_c\cap U_d~.
\end{aligned}
\end{equation}
The first equation is a `categorification' of \eqref{eq:transition_function_principal_1_bundle}: the original equation holds only up to the isomorphism $\sft(h_{abc})$. The second equation is the appropriate non-Abelian generalisation of the defining relation of a \v Cech 2-cocycle.

Clearly, if $\sH=\{\unit\}$, that is, $\sH$ is the trivial group, then the definition \eqref{eq:transition_function_principal_2_bundle} reduces to that of an ordinary principal $\sG$-bundle. Moreover, for the Lie crossed module $\sB\sU(1):=(\sU(1)\stackrel{\sft}{\to}\{ \unit\},\acton)$ with $\sft$ and $\acton$ trivial, this definition coincides with that of an Abelian (bundle) gerbe. Finally, twisted principal bundles can also be regarded as certain principal 2-bundles. We would like to emphasise that, roughly speaking, the cohomology $H^2(M,\sH\rightarrow\sG)$ combines ordinary first-order and second-order \v Cech cohomologies non-trivially: if both $\sH$ and $\sG$ are Abelian and the action of $\sG$ onto $\sH$ is trivial, then $H^2(M,\sH\rightarrow\sG)\cong H^1(M,\sG)\oplus H^2(M,\sH)$. 

Two principal 2-bundles, represented by the transition functions $(\{g_{ab}\},\{h_{abc}\})$ and $(\{\tilde{g}_{ab}\},\{\tilde{h}_{abc}\})$, are considered equivalent (or cohomologous), $(\{g_{ab},h_{abc}\})\sim(\{\tilde g_{ab},\tilde h_{abc}\})$, if there are smooth maps $g_a:U_a\rightarrow \sG$ and $h_{ab}:U_a\cap U_b\rightarrow \sH$ such that
\begin{equation}\label{eq:2BundEquiv}
 g_a \tilde{g}_{ab}\ =\ t(h_{ab})g_{ab}g_b\eand h_{ac}h_{abc}\ =\ (g_a\acton \tilde{h}_{abc})h_{ab}(g_{ab}\acton h_{bc})~.
\end{equation}
A {\it trivial principal 2-bundle} is then described by transition functions that are all cohomologous to one, that is, $g_{ab}\sim\unit$ on all $U_a\cap U_b$ and $h_{abc}\sim\unit$ on all $U_a\cap U_b\cap U_c$.

Note that by virtue of \eqref{eq:2BundEquiv}, we can always assume that $h_{aaa}=\unit$ without loss of generality. Concretely, starting from a general $h_{abc}$ with $h_{aaa}\neq\unit$, it is a straightforward exercise to show that for the choice $h_{aa}=h_{aaa}$ we obtain $\tilde h_{aaa}=\unit$.  In the following, we shall therefore always make this choice and assume $h_{aaa}=\unit$. Clearly, the residual equivalence relations are then those relations \eqref{eq:2BundEquiv} for which $h_{aa}=\unit$. Furthermore, from \eqref{eq:transition_function_principal_2_bundle} together with $h_{aaa}=\unit$ we immediately conclude that also $g_{aa}=\unit$ and $h_{aab}=h_{abb}=\unit$.

\subsection{Cocycle description of principal 3-bundles}

\paragraph{General 3-cocyles.} Let us now move on and discuss smooth principal 3-bundles. There are two obvious ways of categorifying Lie crossed modules and therefore two routes to principal 3-bundles. The first one, using `crossed modules of crossed modules' also known as crossed squares, yields the 2-gerbes of Breen \cite{breen-2006,Breen:1994aa}. The second one, which we shall be following here, was developed by Jur\v co in  \cite{Jurco:2009px} and uses 2-crossed modules $(\sL\overset{\sft}{\to}\sH\overset{\sft}{\to}\sG,\acton,\{\cdot,\cdot\})$. It leads to a nice categorification of the cocycle description \eqref{eq:transition_function_principal_2_bundle} in terms of a specific generalised, non-Abelian \v Cech cohomology, denoted by $H^3(M,\sL\rightarrow \sH\rightarrow \sG)$ in the following.  Representatives of elements of this non-Abelian cohomology  set are 3-cocycles which are collections $(\{g_{ab}\},\{h_{abc}\},\{\ell_{abcd}\})$ of smooth maps $g_{ab}:U_a\cap U_b\to \sG$, $h_{abc}:U_a\cap U_b\cap U_c\to \sH$, and $\ell_{abcd}:U_a\cap U_b\cap U_c\cap U_d\to \sL$ subject to the following cocycle conditions \cite{Jurco:2009px}\footnote{The convention in comparison to \cite{Jurco:2009px} are as follows: $n_{ij}\leftrightarrow g_{ab}$, $m_{ijk}\leftrightarrow h^{-1}_{abc}$ and $\ell_{ijkl}\leftrightarrow \ell_{abcd}$.}: 
\begin{subequations}\label{eq:CoCycle3Bun}
\begin{eqnarray}
 \sft(h_{abc})g_{ab}g_{bc}\! &=&\! g_{ac}\quad\mbox{on}\quad U_a\cap U_b\cap U_c~,\label{eq:CoCycle3Bun-A}\\
 h_{acd}h_{abc}\sft(\ell_{abcd})\! &=&\! h_{abd}(g_{ab}\acton h_{bcd})\quad\mbox{on}\quad U_a\cap U_b\cap U_c\cap U_d~,\label{eq:CoCycle3Bun-B}
\end{eqnarray}
and
\begin{eqnarray}
 \ell_{abcd}((g_{ab}\acton h^{-1}_{bcd})\acton\ell_{abde})(g_{ab}\acton\ell_{bcde})\! &=&\!\notag \\
&&\kern-5cm=\ (h^{-1}_{abc}\acton\ell_{acde})\{h^{-1}_{abc},g_{ac}\acton h^{-1}_{cde}\}((g_{ab}g_{bc}\acton h^{-1}_{cde})\acton \ell_{abce})~.\label{eq:CoCycle3Bun-C}
\end{eqnarray}
\end{subequations}
on $U_a\cap U_b\cap U_c\cap U_d\cap U_e$. The first equation is the same as that for a principal 2-bundle. The next equation is a `categorification' of the corresponding equation of a principal 2-bundle. The last equation is the appropriate non-Abelian generalisation of the Abelian \v Cech cocycle equation; see \cite{Jurco:2009px} for details on the derivation. As before, we will refer to the 3-cocycles as transition functions. 

From the above equations it is clear that the cohomology $H^3(M,\sL\rightarrow\sH\rightarrow\sG)$ combines ordinary first-order, second-order, and third-order \v Cech cohomologies in a non-trivial fashion: if $\sL$, $\sH$, and $\sG$ are all Abelian and the action of $\sG$ onto $\sH$ and $\sL$ is trivial, then $H^3(M,\sL\rightarrow\sH\rightarrow\sG)\cong H^1(M,\sG)\oplus H^2(M,\sH)\oplus H^3(M,\sL)$. Note that these principal 3-bundles are a non-Abelian generalisation of principal 2-bundles that contain Abelian 2-gerbes: for $\sL=\{\unit\}$, the cocycle conditions \eqref{eq:CoCycle3Bun} reduce to the ones of a principal 2-bundle. Moreover, for the Lie 2-crossed module $\sB\sB\sU(1)=(\sU(1)\stackrel{\sft}{\to}\{\unit\}\stackrel{\sft}{\to}\{\unit\})$, we recover the definition of a local Abelian (bundle) 2-gerbe. Note that twisted gerbes are special cases of principal 3-bundles.

Two principal 3-bundles described by transition functions ($\{g_{ab}\},\{h_{abc}\},\{\ell_{abcd}\})$ and $(\{\tilde g_{ab}\},\{\tilde h_{abc}\},\{\tilde\ell_{abcd}\})$ are considered equivalent, if there are smooth maps $g_a:U_a\rightarrow \sG$, $h_{ab}:U_a\cap U_b\rightarrow \sH$ and $\ell_{abc}:U_a\cap U_b\cap U_c\rightarrow \sL$ such that
\begin{subequations}\label{eq:CoBoundary3Bun}
\begin{eqnarray}
  g_a \tilde{g}_{ab}\! &=&\! t(h_{ab})g_{ab}g_b~,\\
  h_{ac}h_{abc}\! &=&\! (g_a\acton \tilde{h}_{abc})h_{ab}(g_{ab}\acton h_{bc})\sft(\ell_{abc})~,\\
  \ell_{abcd}\! &=&\! (h^{-1}_{abc}\acton\ell_{acd})(h_{abc}^{-1}\acton(g_{ac}\acton h_{cd}^{-1}))\acton\notag\\
     &&\kern.5cm\acton\left[\ell_{abc}(((g_{ab}\acton h_{bc})h_{ab})\acton g_a\acton \tilde\ell_{abcd})(h_{ab}^{-1}\acton\{h_{ab},g_{b}\acton h_{bc}^{-1}\})\right]\times\notag\\
  &&\kern.5cm\times\, (h_{ab}^{-1}\acton\{h_{ab},(h_{abc}^{-1}\acton g_{ac}\acton h_{cd}^{-1})\times\notag\\
  &&\kern.5cm\times\, ((\sft(h_{abc}^{-1})g_{ac})\acton h_{cd}^{-1})(h_{abc}^{-1}\acton g_{ac}\acton h_{cd}^{-1})\})\{h^{-1}_{abc},g_{ac}\acton h^{-1}_{cd}\}\times\notag\\
  &&\kern.5cm\times\, (g_{ab}\acton\ell^{-1}_{bcd})(h_{ab}^{-1}\acton\{h_{ab},g_{ab}\acton\sft(\ell_{bcd})\})\times\notag\\
  &&\kern.5cm\times\,((g_{ab}\acton h_{bcd}^{-1})h_{ab}^{-1}\acton\{h_{ab},g_{ab}\acton h_{bcd}\})\times\notag\\
  &&\kern.5cm\times\,((g_{ab}\acton h_{bcd}^{-1}h_{bd}^{-1})h_{ab}^{-1}\acton\{h_{ab},g_{ab}\acton h_{bd}\})((g_{ab}\acton h^{-1}_{bcd})\acton\ell^{-1}_{abd})~.
\end{eqnarray}
\end{subequations}

Similarly to the principal 2-bundle case, these equivalence relations can be used to normalise the transition functions according to
\begin{equation}\label{eq:CocycleNormal}
 g_{aa}\ =\ \unit~,\quad h_{aab}\ =\ h_{abb}\ =\ \unit~,\eand\ell_{aabc}\ =\ \ell_{abbc}\ =\ \ell_{abcc}\ =\ \unit~,
\end{equation}
and we shall always do so in the following.

\paragraph{Trivial 3-cocyles.}
For a {\it trivial principal 3-bundle}, there exist smooth maps $g_a:U_a\to\sG$, $h_{ab}:U_a\cap U_b\to\sH$, and $\ell_{abc}:U_a\cap U_b\cap U_c\to\sL$ such that the transition functions satisfy the following conditions:
\begin{subequations}\label{eq:TrivCoCycle3Bun}
\begin{eqnarray}
 g_{ab}\! &=&\! \sft(h_{ab}^{-1})g_ag_b^{-1}~,\label{eq:TrivCoCycle3Bun-A}\\
 h_{abc}\! &=&\! h_{ac}^{-1}h_{ab}(g_{ab}\acton h_{bc})\sft(\ell_{abc}^{-1})~,\label{eq:TrivCoCycle3Bun-B}\\
\ell_{abcd}\! &=&\! (h^{-1}_{abc}\acton\ell_{acd})\{h^{-1}_{abc},g_{ac}\acton h^{-1}_{cd}\}\times\notag\\
  &&\kern1cm \times\,((g_{ab}g_{bc}\acton h^{-1}_{cd})\acton \ell_{abc})(g_{ab}\acton\ell^{-1}_{bcd})((g_{ab}\acton h^{-1}_{bcd})\acton\ell^{-1}_{abd})~.\label{eq:TrivCoCycle3Bun-C} 
\end{eqnarray}
\end{subequations}
These relations can be derived from \eqref{eq:CoBoundary3Bun} by putting $\tilde{g}_{ab}=\unit$, $\tilde{h}_{abc}=\unit$, and $\tilde{\ell}_{abc}=\unit$. Alternatively, the first of these equations can be read off \eqref{eq:CoCycle3Bun-A} by fixing the patch where the cocycle trivialises to be $U_c$. Likewise, \eqref{eq:TrivCoCycle3Bun-B} follows from \eqref{eq:CoCycle3Bun-B} by fixing the trivialising patch to be $U_d$ and \eqref{eq:TrivCoCycle3Bun-C} follows from \eqref{eq:CoCycle3Bun-C} by fixing the trivialising patch to be $U_e$, respectively.  Note that to preserve the normalisations \eqref{eq:CocycleNormal}, we also normalise the group-valued functions $h_{ab}$ and $\ell_{abc}$ according to
\begin{equation}\label{eq:SplittingNormal}
 h_{aa}\ =\ \unit\eand \ell_{aaa}\ =\ \ell_{aab}\ =\ \ell_{abb}\ =\ \unit~.
\end{equation}
Finding the splitting \eqref{eq:TrivCoCycle3Bun} for a given collection of transition functions $(\{g_{ab}\},\{h_{abc}\},\{\ell_{abcd}\})$ amounts to solving a Riemann--Hilbert problem.

\section{Penrose--Ward transform and self-dual fields}\label{sec:PWTSDF}

We are now in the position to pursue our main idea to construct self-dual non-Abelian tensor field theories in six dimensions via twistor theory. The basic idea of twistor theory is to encode solutions to certain field equations on space-time in terms of holomorphic data on a twistor space that is associated with space-time in a particular fashion. Conversely, the field equations in question and the associated gauge transformations follow naturally from the {\it algebraic} twistor data. For gauge theory, this map between data on space-time and data on twistor space is known as a {\em Penrose--Ward transform}. 

The twistor approach has been used very successfully for the description of instantons in Yang--Mills theory \cite{Ward:1977ta,Atiyah:1977pw} and Einstein gravity \cite{Penrose:1976js,Atiyah:1978wi,Ward:1980am} in four dimensions (including their supersymmetric extensions, e.g.~in \cite{Manin:1988ds,Merkulov:1991kt,Merkulov:1992,Merkulov:1992qa,Merkulov:1992b,Witten:2003nn,Popov:2004rb,Berkovits:0406051,Wolf:2007tx,Mason:2007ct}). In this context, the twistor space is the Penrose twistor space \cite{Penrose:1967wn,Penrose:1968me,Penrose:1969aa,Penrose:1972ia} (and supersymmetric extensions thereof, see for example \cite{Ferber:1977qx,Manin:1988ds,Witten:2003nn}) and the holomorphic twistor data are certain holomorphic vector and principal bundles in the case of Yang--Mills theory and holomorphic contact structures in the case of Einstein gravity. The twistor approach has also been used for description of gauge theory instantons in dimensions greater than four (see, e.g.~\cite{Ward:1983zm,Popov:2009nx,Wolf:2009ep,Wolf:2012gz,Lechtenfeld:2012ye,Ivanova:2013vya} and references therein).  In addition, it has been used to describe equations of motion of non-self-dual theories such as maximally supersymmetric Yang--Mills theory in four \cite{Witten:1978xx,Isenberg:1978kk,Buchdahl:1985aa,Manin:1988ds}, three \cite{Saemann:2005ji}, and six dimensions \cite{Saemann:2012rr}. 

More recently, it was demonstrated that twistor theory can also be applied to the construction of superconformal non-Abelian self-dual tensor field theories in six dimensions \cite{Saemann:2012uq} which are based on principal 2-bundles (see \cite{Saemann:2011nb,Mason:2011nw,0198535651,Mason:2012va} for the Abelian description). In the remainder of this work, we would like to generalise our previous results \cite{Saemann:2012uq} by constructing superconformal self-dual non-Abelian tensor field theories in six dimensions that are based on principal 3-bundles. As indicated previously, principal 3-bundles allow to relax the Peiffer identity, and, as such, we will obtain self-dual tensor theories with a 3-form curvature that is significantly less constrained than that of principal 2-bundles.

\subsection{Outline of the Penrose--Ward transform}

Since the following construction is rather lengthy and technical, let us give a brief overview over the key components and present an outline of the Penrose--Ward transform discussed in this section. We are interested in an $\CN=(n,0)$ superconformal gauge theory in six dimensions, and we start by constructing a twistor space $P^{6|8n}$ to six-dimensional flat Minkowski superspace $M^{6|8n}$. Relating both spaces is the so-called correspondence space $F^{9|8n}$, which is fibred over both $P^{6|8n}$ and $M^{6|8n}$:
\begin{equation}\label{eq:superDoubleFibration}
 \begin{picture}(50,40)
  \put(0.0,0.0){\makebox(0,0)[c]{$P^{6|2n}$}}
  \put(64.0,0.0){\makebox(0,0)[c]{$M^{6|8n}$}}
  \put(34.0,33.0){\makebox(0,0)[c]{$F^{9|8n}$}}
  \put(7.0,18.0){\makebox(0,0)[c]{$\pi_1$}}
  \put(55.0,18.0){\makebox(0,0)[c]{$\pi_2$}}
  \put(25.0,25.0){\vector(-1,-1){18}}
  \put(37.0,25.0){\vector(1,-1){18}}
 \end{picture}
\end{equation}
The aim of the Penrose--Ward transform is now to establish a bijection between certain topologically trivial holomorphic principal 3-bundles over twistor space $P^{6|2n}$ and solutions to the superconformal gauge theory over $M^{6|8n}$. Starting from such a principal 3-bundle $\hat{E}\rightarrow P^{6|2n}$, we pull it back along $\pi_1$. The resulting principal 3-bundle $E$ on correspondence space $F^{9|8n}$ turns out to be holomorphically trivial. There exists now a differential cohomology relative to $\pi_1$, which translates between a holomorphically trivial principal 3-bundle with non-trivial transition functions but vanishing relative connective structure and a principal 3-bundle with trivial transition functions and flat but non-vanishing relative connective structure. The explicit construction of this equivalence relation will turn out to be the most involved part of our construction. The resulting flat relative connective structure on correspondence space can be readily pushed down to $M^{6|8n}$. Relative flatness translates into certain constraint equations which are fulfilled by the connective structure on $M^{6|8n}$. These constraint equations are equivalent to the field equations of the superconformal field theory we are interested in. Besides the constraint equations, our construction also exposes the full gauge symmetry of connective structures on principal 3-bundles.

\subsection{Twistor space}

\paragraph{Chiral superspace.}
As we wish to discuss superconformal gauge theories in six dimensions, let us consider {\it complexified} flat space-time, $M^6:=\FC^6$, and extend it by $8n$ fermionic directions with $n\in\{0,1,2\}$. We obtain $\CN=(n,0)$ chiral superspace $M^{6|8n}:=\FC^{6|8n}:=\FC^6\oplus\Pi\FC^{8n}$, where $\Pi$ is the (Gra{\ss}mann) parity-changing operator. We will always work in this complexified setting; real structures leading to Minkowski signature or split signature can be introduced whenever desired (see, e.g., \cite{Saemann:2011nb} for details). 

Next, we coordinatise chiral superspace $M^{6|8n}$ by  $(x^{AB},\eta^A_I)$, where the $x^{AB}=-x^{BA}$ are bosonic (i.e.~Gra{\ss}mann-parity even) coordinates and the $\eta^A_I$ are fermionic (i.e.~Gra{\ss}\-mann-parity odd) coordinates. The indices $A,B,\ldots=1,\ldots,4$ are (anti-chiral) spinor indices and $I,J,\ldots=1,\ldots,2n$ are $\sSp(n)$ R-symmetry indices. Anti-symmetric pairs of spinor indices may be raised and lowered with the help of the Levi-Civita symbol $\frac12\varepsilon_{ABCD}$. Specifically, we shall write $x_{AB}:= \frac12\varepsilon_{ABCD}x^{CD}$.

If we let $\Omega:=(\Omega^{IJ})$ be an $\sSp(n)$-invariant $2n\times 2n$ matrix, we may introduce the derivative operators
\begin{equation}\label{eq:vector_fields_space_time}
 P_{AB}\ :=\ \der{x^{AB}}\ :=\ \partial_{AB} \eand D^I_A\ :=\ \der{\eta^A_I}-2\Omega^{IJ}\eta_J^B\der{x^{AB}}~. 
\end{equation}
They obey the (anti-)commutation relations
\begin{equation}
 [P^{AB},P^{CD}]\ =\ 0~,\quad
 [P^{AB},D^I_A] =\ 0~,\eand [D^I_A,D^J_B]\ =\ -4\Omega^{IJ}P_{AB}~,
\end{equation}
where $[\cdot,\cdot]$ denotes the super Lie bracket (graded supercommutator).

In the following, we will use the conventions $\partial_{AB}x^{CD}=\delta_{[A}^{[C}\delta_{B]}^{D]}$, where brackets denote normalised anti-symmetrisation of the enclosed indices.\footnote{Likewise, parentheses will denote normalised symmetrisation of the enclosed indices.}  

\paragraph{Twistor space and double fibration.}
To define the relevant twistor space of  $M^{6|8n}$, we first introduce the {\it correspondence space}, which we define as $F^{9|8n}:=M^{6|8n}\times\PP^3$ with $\PP^3$ being complex projective 3-space. Furthermore, we equip the correspondence space with the coordinates $(x^{AB},\eta^A_I,\lambda_A)$, where the $\lambda_A$ are homogeneous coordinates\footnote{In particular, all equations involving these homogeneous coordinates are obviously to be understood as homogeneous equations.} on $\PP^3$. Notice that for $n=0$, the correspondence space can be understood as the projectivisation of  the dual of the (rank-4) bundle of anti-chiral spinors on six-dimensional space-time $M^6$.

On $F^{9|8n}$, we introduce a distribution $D:= \mbox{span}\{V^A, V^{IAB}\}$, called the {\it twistor distribution}, with
\begin{equation}
  V^A\ :=\ \lambda_B\partial^{AB}\eand
  V^{IAB}\ :=\ \tfrac12\varepsilon^{ABCD}\lambda_C D^I_D~.
\end{equation}
This is a rank-$3|6n$ distribution because of the relations $\lambda_A V^A=0=\lambda_B V^{IAB}$. Moreover, it is a straightforward exercise to check that $D$ is an integrable distribution, that is, $[D,D]\subseteq D$. Therefore, we have a foliation of the correspondence space, and we define the twistor space to be the quotient $P^{6|2n}:=F^{9|8n}/D$. This is a complex (super)manifold. Indeed, if we let $(z^A,\lambda_A,\eta_I)$ be homogeneous coordinates on $\PP^{7|2n}$, then the twistor space $P^{6|2n}$ is a complex hypersurface in $\PP^{7|2n}\setminus\PP^{3|2n}$ given by the zero locus
\begin{equation}\label{eq:superquadric}
 z^A\lambda_A-\Omega^{IJ}\eta_I\eta_J\ =\ 0~,
\end{equation}
where the $\PP^{3|2n}$ we remove from  $\PP^{7|2n}$ is given by the condition $\lambda_A=0$.

In summary, we have established the double fibration \eqref{eq:superDoubleFibration}, where $\pi_2$ denotes the trivial projection while $\pi_1$ acts as
\begin{equation}
\pi_1:(x^{AB},\eta^A_I,\lambda_A)\ \mapsto\ (z^A,\eta_I,\lambda_A)\ =\ ((x^{AB}+\Omega^{IJ}\eta^A_I\eta^B_J)\lambda_B~,~\eta_I^A\lambda_A~,~\lambda_A)~.
\end{equation}
The equations 
\begin{equation}\label{eq:superincidence}
 z^A\ =\ (x^{AB}+\Omega^{IJ}\eta^A_I\eta^B_J)\lambda_B\eand
 \eta_I\ =\ \eta_I^A\lambda_A
\end{equation}
are called the {\it incidence relation}.

\paragraph{Geometric twistor correspondence.}
Because of the incidence relation \eqref{eq:superincidence}, we have a geometric relation between points and certain submanifolds. Specifically, any point $x\in M^{6|8n}$ in chiral superspace corresponds to a complex projective 3-space  $\hat x=\pi_1(\pi_2^{-1}(x))\hookrightarrow P^{6|2n}$. Conversely, any point $p\in P^{6|2n}$ in twistor space corresponds to a $3|6n$-superplane $\pi_2(\pi_1^{-1}(p))\hookrightarrow M^{6|8n}$,
\begin{equation}\label{eq:IncidenceSolutionSuper}
\begin{aligned}
 x^{AB}\ &=\ x_0^{AB}+\varepsilon^{ABCD}\mu_C\lambda_D+2\Omega^{IJ}\varepsilon^{CDE[A}\lambda_C\theta_{IDE}\eta_0{}^{B]}_J~,\\
 \eta^A_I\ &=\ \eta_0{}^A_I+\varepsilon^{ABCD}\lambda_B\theta_{ICD}~.
 \end{aligned}
\end{equation}
Here, $(x_0^{AB},\eta_0{}^A_I)$ is a particular solution to the incidence relation \eqref{eq:superincidence} which corresponds to a reference point in the superplane in $M^{6|8n}$. The moduli $\mu_A$ and $\theta_{IAB}$ are defined up to terms proportional to $\lambda_A$. As a result, there are $3|6n$ (which is the rank of the twistor distribution) moduli parametrising the superplane. Note that for $n=0$, \eqref{eq:IncidenceSolutionSuper} reduce to $ x^{AB}=x_0^{AB}+\varepsilon^{ABCD}\mu_C\lambda_D$ and $\varepsilon^{ABCD}\mu_C\lambda_D$ is a generic null-vector in six dimensions. Hence, \eqref{eq:IncidenceSolutionSuper} represent a super-extension of a totally null 3-plane. 

\subsection{Penrose--Ward transform}

\paragraph{Twistor data.}
Subject of this section is the explicit derivation of non-Abelian self-dual tensor theories by means of twistor theory. In particular, the algebraic twistor data from which we would like to start is represented by topologically trivial, {\it holomorphic} principal 3-bundles $\hat E\to P^{6|2n}$ over the  twistor space $P^{6|2n}$ which, in addition, are holomorphically trivial when restricted to any complex projective 3-space $\hat x=\pi_1(\pi_2^{-1}(x))\hookrightarrow P^{6|2n}$. Following Manin's terminology in the principal 1-bundle case \cite{Manin:1988ds}, we will refer to such bundles as {\em $M^{6|8n}$-trivial}. We denote the structure 3-group of $\hat E$ by $(\sL\overset{\sft}{\to}\sH\overset{\sft}{\to}\sG,\acton,\{\cdot,\cdot\})$.\footnote{At this stage, we could assume that $\sL$, $\sH$, and $\sG$ are Lie supergroups, and hence, $\hat E$ is a principal 3-superbundle. However, for the sake of clarity, and since there is no immediate physical application of supergroups here, we will only work with Lie groups. The discussion below can be adapted to the supergroup case without any difficulty.} If, as before, $\{\hat U_a\}$ is an open covering of $P^{6|2n}$ that is chosen to be sufficiently fine (Stein), then $\hat E\to P^{6|2n}$ can be described by holomorphic transition functions  $(\{\hat g_{ab}\},\{\hat h_{abc}\},\{\hat \ell_{abcd}\})$ on appropriate non-empty overlaps of the coordinate patches $\hat U_a$ subject to the cocycle conditions \eqref{eq:CoCycle3Bun} and the normalisations \eqref{eq:CocycleNormal}.\footnote{The standard (Stein) cover of $P^{6|2n}$ consists of four coordinate patches. Hence, for this choice of cover, there will only be one $\ell_{abcd}$, and, consequently, the cocycle condition \eqref{eq:CoCycle3Bun-C} for $\ell_{abcd}$ will become vacuous.} 

\paragraph{Relative connective structure.}
Next, we wish to pull back such an $M^{6|8n}$-trivial bundle $\hat E\to P^{6|2n}$ to the correspondence space along the projection $\pi_1$. This yields a holomorphic principal 3-bundle $E:=\pi_1^*\hat E$ over $F^{9|8n}$. For simplicity, we work with the induced covering $\{U_a:=\pi_1^{-1}(\hat U_a)\}$ on the correspondence space. This enables us to describe the bundle $E$ in terms of the pulled-back transition functions $(\{g_{ab}:=\pi_1^*\hat g_{ab}\},\{h_{abc}:=\pi_1^*\hat h_{abc}\},\{\ell_{abcd}:=\pi_1^*\hat \ell_{abcd}\})$ which also obey \eqref{eq:CocycleNormal}. Since we are assuming that the bundle $\hat E\to P^{6|2n}$ is topologically trivial, and, in addition, holomorphically trivial on all projective 3-spaces $\hat x=\pi_1(\pi_2^{-1}(x))$, the bundle $E\to F^{9|8n}$ must be trivial (topologically as well as holomorphically) on all of the correspondence space. Hence, by virtue of \eqref{eq:TrivCoCycle3Bun}, its transition functions are of the form
\begin{subequations}\label{eq:HolSplitCorr}
\begin{eqnarray}
 g_{ab}\! &=&\! \sft(h_{ab}^{-1})g_ag_b^{-1}~,\label{eq:PullBack-A}\\
 h_{abc}\! &=&\! h_{ac}^{-1}h_{ab}(g_{ab}\acton h_{bc})\sft(\ell_{abc}^{-1})~,\label{eq:PullBack-B}\\
\ell_{abcd}\! &=&\! (h^{-1}_{abc}\acton\ell_{acd})\{h^{-1}_{abc},g_{ac}\acton h^{-1}_{cd}\}\times\notag\\
  &&\kern1cm \times\,((g_{ab}g_{bc}\acton h^{-1}_{cd})\acton \ell_{abc})(g_{ab}\acton\ell^{-1}_{bcd})((g_{ab}\acton h^{-1}_{bcd})\acton\ell^{-1}_{abd})~.\label{eq:PullBack-C} 
\end{eqnarray}
\end{subequations}
We emphasise that the group-valued functions  $(\{ g_{a}\},\{ h_{ab}\},\{\ell_{abc}\})$ are holomorphic and obey \eqref{eq:SplittingNormal}. Despite being holomorphically trivial, the bundle  $E\to F^{9|8n}$ contains non-trivial information: the explicit equivalence relation turning the transition functions describing $E$ to trivial ones generate a non-trivial {\it relative connective structure} as will become transparent momentarily. 

By definition of a pull-back, the transition functions $(\{ g_{ab}\},\{ h_{abc}\},\{\ell_{abce}\})$ of $E$ must be annihilated by the twistor distribution. Specifically, let us introduce the relative exterior derivative along the fibration $\pi_1:  F^{9|8n}\to  P^{6|2n}$ by
\begin{equation}\label{eq:DefOfRelDer}
  \dd_{\pi_1}\ :=\ e_A V^A+e_{IAB} V^{IAB}\ =\ e_{[A}\lambda_{B]}\partial^{AB}+e^A_I D^I_A~,
\end{equation}
where we have defined $e_I^A:=-\frac12\varepsilon^{ABCD}e_{IBC}\lambda_D$.  The 1-forms $e_A$ and $e_{IAB}$ span the dual of the twistor distribution, and they are of homogeneity $-1$ and defined up to terms proportional to $\lambda_A$ because of the relations $\lambda_A V^A=0=\lambda_B V^{IAB}$. As an immediate consequence of the (anti-)commutation relations of the twistor distribution, Cartan's structural equations for the 1-forms $e_A$ and $e^I_A$ read as
\begin{equation}\label{eq:MCSE}
  \dd_{\pi_1}(e_{[A}\lambda_{B]})\ =\ \Omega^{IJ}\varepsilon_{ABCD} e^C_I\wedge e^D_J\eand
  \dd_{\pi_1}e^A_I\ =\ 0~.
\end{equation}
Thus, we have
\begin{equation}
 \dd_{\pi_1}g_{ab}\ =\ \dd_{\pi_1}h_{abc}\ =\ \dd_{\pi_1}\ell_{abcd}\ =\ 0
 \end{equation}
but, importantly, $\dd_{\pi_1}g_{a}\neq0$, $\dd_{\pi_1}h_{ab}\neq0$, and $\dd_{\pi_1}\ell_{abc}\neq0$, in general. This latter fact enables us to introduce a set of differential 1-forms
\begin{subequations}
\begin{eqnarray}
 \kern-.5cm a_a\! &:=&\! g_a^{-1}\dd_{\pi_1} g_a~,\\
 \kern-.5cm b_{ab}\! &:=&\!g_a^{-1}\acton(\dd_{\pi_1} h_{ab}h_{ab}^{-1})~,\\
 \kern-.5cm c_{abc}\! &:=&\! g_a^{-1}\acton\big[(h_{ab}(g_{ab}\acton h_{bc}))\acton (\ell^{-1}_{abc}\dd_{\pi_1}\ell_{abc})-\{h_{ab},(\sft(h_{ab}^{-1})g_a)\acton b_{bc}\}\big]~,
\end{eqnarray}
\end{subequations}
on the appropriate intersections of the coordinate patches $\{U_a\}$. Notice that these forms have components only along the fibration $\pi_1:  F^{9|8n}\to  P^{6|2n}$, and, as such, they are so-called {\em relative differential 1-forms}. In our subsequent discussion, we will denote the sheaf of relative differential $r$-forms on the correspondence space by $\Omega^r_{\pi_1}$.  Notice also that $b_{aa}=c_{aaa}=c_{aab}=c_{abb}=0$ because of the normalisations \eqref{eq:SplittingNormal} and the fact that $\{\unit,Y\}=0$ for all $Y\in\frh$. 

The relative 1-forms $a_a$, $b_{ab}$, and $c_{abc}$ obey the following relations:
\begin{subequations}
\begin{eqnarray}
  a_a \! &=&\! a_b+\sft(b_{ab})\quad\mbox{on}\quad U_a\cap U_b~,\label{eq:CocycleCond-a}\\
  b_{ab}+b_{bc} \! &=&\! b_{ac}+\sft(c_{abc})\quad\mbox{on}\quad U_a\cap U_b\cap U_c~,\label{eq:CocycleCond-b}\\
  c_{abc}-c_{bcd}+c_{cda}\! &=&\! c_{dab}\quad\mbox{on}\quad U_a\cap U_b\cap U_c\cap U_d~.\label{eq:CocycleCond-c}
\end{eqnarray}
\end{subequations}
The first two of these equations follow relatively straightforwardly from \eqref{eq:PullBack-A} and \eqref{eq:PullBack-B}, respectively, while the last equation follows from \eqref{eq:PullBack-C} after a lengthy calculation. In fact, we have established the equation \eqref{eq:CocycleCond-c} and the more involved relations following below using a computer algebra programme.\footnote{The procedure we have used in deriving this result is as follows: i) simplify the equation to $c_{abc}-c_{bcd}+c_{cda}=c_{dab}+\ldots $ with a minimum amount of substitutions of the expressions for trivial cocycles; ii)
 turn all actions of $\sH$ on $\frl$ in the remaining terms into actions of $\sG$ on $\frl$; iii) simplify all the terms by using the identity $ \{h_1,h_1^{-1} \acton h_2\}=\{h_1,h_1^{-1}h_2h_1\}=-\sft(h_1)\acton\{h_1^{-1},h_2\}$ for $h_1,h_2\in\sH$.} 
 
Next, we point out that equation \eqref{eq:CocycleCond-c} implies that $c_{abc}$ is totally anti-symmetric in its indices:  since $c_{aaa}=c_{aab}=c_{baa}=0$, the choice $a=c$ in \eqref{eq:CocycleCond-c} yields $c_{aba}=0$. Using this, we further  obtain for $a=b$, $b=c$, and $c=d$ in   \eqref{eq:CocycleCond-c} the equations $c_{abc}=c_{bca}$, $c_{abc}=-c_{cba}$, and $c_{abc}=c_{cab}$, respectively, which together imply the total anti-symmetry of $c_{abc}$. Therefore, we may conclude that the collection $\{c_{abc}\}$ defines an ordinary $\frl$-valued \v Cech cocycle, that is, it represents an element of the Abelian \v Cech cohomology group $H^2(F^{9|8n},\Omega^1_{\pi_1}\otimes\frl)$. This cohomology group, however, vanishes since $H^2(F^{9|8n},\Omega^1_{\pi_1})$ is zero as follows from similar arguments as those given in 
\cite{0198535651,Saemann:2011nb}. Hence, $c_{abc}$ must be of the form
\begin{equation}\label{eq:Split-c}
 c_{abc}\ =\ c_{ab}+c_{bc}+c_{ca}\ewith c_{ab}\ =\ -c_{ba}~,
\end{equation}
where $c_{ab}$ is defined on $U_a\cap U_b$.

Since $c_{abc}$ is totally anti-symmetric, we conclude that also $b_{ab}$ must be anti-symmetric as this follows from \eqref{eq:CocycleCond-b}  for the choice $c=a$. Upon substituting the splitting \eqref{eq:Split-c} into  \eqref{eq:CocycleCond-b}, we obtain
\begin{equation} \label{eq:RealCocycleCond-b}
 b'_{ab}+b'_{bc}\ =\ b'_{ac}\ewith b'_{ab}\ :=\ b_{ab}-\sft(c_{ab})~.
\end{equation}
Thus, the collection $\{b'_{ab}\}$ defines an element of  the Abelian \v Cech cohomology group $H^1(F^{9|8n},\Omega^1_{\pi_1}\otimes\frh)$. This cohomology group is also zero \cite{0198535651,Saemann:2011nb} and therefore, we have yet another splitting
\begin{equation}\label{eq:Split-b'}
 b'_{ab}\ =\ b'_a-b'_b~.
\end{equation}

Next, we substitute the definition \eqref{eq:RealCocycleCond-b} of $b'_{ab}$ into \eqref{eq:CocycleCond-a} and because of $\sft^2(Z)=0$ for all $Z\in\frl$ and \eqref{eq:Split-b'}, we realise that
\begin{equation}\label{eq:DefOf-A}
 A_a\ :=\ a_a-\sft(b'_a)
\end{equation}
satisfies $A_a=A_b$ on $U_a\cap U_b$. Hence, we obtain a globally defined $\frg$-valued relative 1-form $A_{\pi_1}$ with $A_a=A_{\pi_1}|_{U_a}$.

A straightforward calculation shows that\footnote{Here and in the following, the bracket $[\cdot,\cdot]$ includes the wedge product of forms: if $\omega=\omega^a\tau_a$ and $\rho=\rho^b\tau_b$ are two Lie-algebra valued differential $r$- and $s$-forms in some basis $\tau_a$ of the Lie algebra under consideration, then $[\omega,\rho]=\omega^a\wedge \rho^b[\tau_a,\tau_b]$. The same holds for other operations in the Lie 2-crossed module, in particular for the Peiffer lifting.}
\begin{equation}\label{eq:SomeEqFor-ab}
 \dd_{\pi_1} a_a+\tfrac12 [a_a,a_a]\ =\ 0\eand \dd_{\pi_1} b_{ab}+\tfrac12(a_a+a_b)\acton b_{ab}-\tfrac12\langle b_{ab},b_{ab}\rangle\ =\ 0~.
\end{equation}
where $\langle b_{ab},b_{ab}\rangle =\sft(\{b_{ab},b_{ab}\})$ as before. With the help of the splitting \eqref{eq:Split-b'} and the definition \eqref{eq:RealCocycleCond-b}, these equations imply that
\begin{subequations}
\begin{equation}\label{eq:PreDefOfB}
 -\dd_{\pi_1} b'_a-a_a\acton b'_a+\tfrac12\sft(b'_a)\acton b'_a\ =\  -\dd_{\pi_1} b'_b-a_b\acton b'_b+\tfrac12\sft(b'_b)\acton b'_b+\sft(c'_{ab})
\end{equation}
with 
\begin{equation}
 c'_{ab}\ :=\ \dd_{\pi_1}c_{ab}+\tfrac12(a_a+a_b)\acton c_{ab}+\{b'_{[a},b'_{b]}\}~.
\end{equation}
\end{subequations}
Again, a lengthy calculation shows that
\begin{equation}
 c'_{ab}+c'_{bc}\ =\ c'_{ac}\quad\mbox{on}\quad U_a\cap U_b\cap U_c~.
\end{equation}
This simply says that the collection $\{c'_{ab}\}$ defines an element of the Abelian \v Cech cohomology group 
$H^1(F^{9|8n},\Omega^2_{\pi_1}\otimes\frl)$. However, also this cohomology group vanishes following the arguments of \cite{0198535651,Saemann:2011nb}. Therefore, we arrive at the splitting
\begin{equation}\label{eq:Split-c'}
 c'_{ab}\ =\ c'_a-c'_b~.
\end{equation}
This result can be substituted into \eqref{eq:PreDefOfB} and we conclude that
\begin{equation}\label{eq:DefOf-B}
  B_a\ :=\ -\dd_{\pi_1} b'_a-a_a\acton b'_a+\tfrac12\sft(b'_a)\acton b'_a-\sft(c'_a)
\end{equation}
obeys $B_a=B_b$ on the overlaps $U_a\cap U_b$. Thus, the $B_a$s yield a globally defined $\frh$-valued relative 2-form $B_{\pi_1}$ with $B_a=B_{\pi_1}|_{U_a}$.

Finally, we define a relative differential 3-form
\begin{equation}\label{eq:DefOf-C}
\begin{aligned}
 C_a\ &:=\ -\dd_{\pi_1}\big(c'_a+\tfrac12\{b'_a,b'_a\}\big)-\big(a_a-\sft(b'_a)\big)\acton \big(c'_a+\tfrac12\{b'_a,b'_a\}\big)\,-\\
    &\kern3cm-\big\{b'_a,\dd_{\pi_1}b'_a+(a_a-\tfrac12 \sft(b'_a))\acton b'_a-\tfrac12\langle b'_a,b'_a\rangle\big\}
 \end{aligned}
\end{equation}
on $U_a$. A lengthy computation using a computer algebra programme shows that this 3-form obeys $C_a=C_b$ on $U_a\cap U_b$. Therefore, we have obtained a globally defined $\frl$-valued relative 3-form $C_{\pi_1}$ with $C_a=C_{\pi_1}|_{U_a}$. 

Summarising, we have constructed the global relative differential forms $A_{\pi_1}$, $B_{\pi_1}$, and $C_{\pi_1}$ using the data \eqref{eq:HolSplitCorr}. We may interpret these forms as the connection forms constituting  a {\it relative connective structure} on the principal 3-bundle $E\to F^{9|8n}$.  

\paragraph{Relative curvatures.}
The set of curvature forms associated with the relative connection forms $A_{\pi_1}$, $B_{\pi_1}$, and $C_{\pi_1}$ will consist of a $\frg$-valued 2-form curvature $F_{\pi_1}$, an $\frh$-valued 3-form curvature $H_{\pi_1}$, and an $\frl$-valued 4-form curvature $G_{\pi_1}$.  In particular, using the first equation in \eqref{eq:SomeEqFor-ab} and the definitions \eqref{eq:DefOf-A}, \eqref{eq:DefOf-B}, and \eqref{eq:DefOf-C}, it immediately follows that
\begin{subequations}\label{eq:DefOfFH}
\begin{eqnarray}
  \sft(B_{\pi_1})\! &=&\! \dd_{\pi_1}A_{\pi_1}+\tfrac12[A_{\pi_1},A_{\pi_1}]\ =:\ F_{\pi_1}~,\label{eq:DefOfF}\\
   \sft(C_{\pi_1})\! &=&\! \dd_{\pi_1}B_{\pi_1}+A_{\pi_1}\acton B_{\pi_1}\ =:\ H_{\pi_1}~,\label{eq:DefOfH}
\end{eqnarray} 
\end{subequations}
which define the curvature forms $F_{\pi_1}$ and $H_{\pi_1}$. These equations simply state the vanishing of the so-called 2-form and 3-form {\it fake curvatures},
\begin{equation}\label{eq:rel-fake-curvatures}
 \CF_{\pi_1}\ :=\ F_{\pi_1}-\sft(B_{\pi_1})\eand
 \CH_{\pi_1}\ :=\ H_{\pi_1}-\sft(C_{\pi_1})~.
\end{equation}

It remains to define the 4-form curvature. It is
\begin{equation}\label{eq:DefOfG}
G_{\pi_1}\ :=\ \dd_{\pi_1} C_{\pi_1}+A_{\pi_1}\acton C_{\pi_1}+\{B_{\pi_1},B_{\pi_1}\}~.
\end{equation}
This choice of $G_{\pi_1}$ is essentially dictated by demanding covariant behaviour under gauge transformations, and we will come back to this point below. One can check that $G_a=G_{\pi_1}|_{U_a}=0$, which follows upon substituting the explicit expressions \eqref{eq:DefOf-A}, \eqref{eq:DefOf-B}, and \eqref{eq:DefOf-C} for $A_a$, $B_a$, and $C_a$ into the definition \eqref{eq:DefOfG}. We shall call connective structures, for which the fake relative curvatures \eqref{eq:rel-fake-curvatures} and $G_{\pi_1}$ vanish {\em relative flat}. Note that in the purely bosonic case $n=0$, \eqref{eq:DefOfG} vanishes trivially as in this case the fibres of $\pi_1:  F^{9|0}\to  P^{6|0}$ are three-dimensional implying that there are no relative 4-forms. However, for $n>0$ this expression is, in general, non-trivial because of the extra fermionic directions. As we shall see below, the equation $G_{\pi_1}=0$ will correspond to certain constraint equations on chiral superspace.

Altogether, $M^{6|8n}$-trivial holomorphic principal 3-bundles over the twistor space correspond to holomorphic principal 3-bundles over the correspondence space that are equipped with a relative connective structure that is relative flat.

\paragraph{Gauge freedom on the correspondence space.}
By construction, it is clear that the solutions to the Riemann--Hilbert problems \eqref{eq:HolSplitCorr}, \eqref{eq:Split-c}, \eqref{eq:Split-b'}, and \eqref{eq:Split-c'} are not unique: we can always consider the transformation $g_a\mapsto g_ag$ for a globally defined holomorphic $\sG$-valued function $g$.\footnote{That is, it is independent of $\lambda_A$.} Likewise, we may consider the shifts $b'_a\mapsto b'_a+\Lambda_{\pi_1}$, $c_{ab}\mapsto c_{ab}-\Theta_a+\Theta_b$,\footnote{Hence, the definition of $b'_{ab}$ in \eqref {eq:RealCocycleCond-b} depends on such $\Theta_a$-shifts.} and $c'_a\mapsto c'_a+\Sigma_{\pi_1}$ where $\Lambda_{\pi_1}$ is a globally defined $\frh$-valued relative 1-form, $\Theta_a$ is an $\frl$-valued relative 1-form defined on $U_a$, and $\Sigma_{\pi_1}$ is a globally defined $\frl$-valued relative 2-form, since these shifted forms represent equally good solutions to the Riemann--Hilbert problems \eqref{eq:Split-c}, \eqref{eq:Split-b'}, and \eqref{eq:Split-c'}. The combination of these transformations then yields
\begin{subequations}\label{eq:GaugeFreedomSmallForms}
\begin{eqnarray}
 a_a\! &\mapsto&\! \tilde a_a\ :=\ g^{-1}a_a g+g^{-1}\dd_{\pi_1}g~,\label{eq:GTa}\\
 b'_a\! &\mapsto&\! \tilde b'_a\ :=\ g^{-1}\acton b'_a+\Lambda_{\pi_1}+\sft(\Theta_a)~,\label{eq:GTb'}\\
 c'_a\! &\mapsto&\! \tilde c'_a\ :=\ g^{-1}\acton c'_a-\dd_{\pi_1}\Theta_a-\tilde a_a\acton\Theta_a+\tfrac12\sft(\tilde b'_a-\Lambda_{\pi_1})\acton\Theta_a+\notag \\
  &&\kern4cm+\,\tfrac12\{\tilde b'_a,\Lambda_{\pi_1}\}-\tfrac12\{\Lambda_{\pi_1},\tilde b'_a\}+\Sigma_{\pi_1}~.\label{eq:GTc'}
\end{eqnarray}
\end{subequations}
Some lengthy algebraic manipulations show that under these transformations, the relative connection forms  \eqref{eq:DefOf-A}, \eqref{eq:DefOf-B}, and \eqref{eq:DefOf-C} behave as
\begin{subequations}\label{eq:GTConRelForms}
\begin{eqnarray}
 \kern-1cm A_{\pi_1}\! &\mapsto&\! \tilde A_{\pi_1}\ :=\ g^{-1}A_{\pi_1} g+g^{-1}\dd_{\pi_1}g-\sft(\Lambda_{\pi_1})~,\label{eq:GTA}\\
 \kern-1cm  B_{\pi_1}\! &\mapsto&\! \tilde B_{\pi_1}\ :=\ g^{-1}\acton B_{\pi_1}-\tilde \nabla_{\pi_1}\Lambda_{\pi_1}-\tfrac12\sft(\Lambda_{\pi_1})\acton \Lambda_{\pi_1}-\sft(\Sigma_{\pi_1})~,\label{eq:GTB}\\
\kern-1cm C_{\pi_1}\! &\mapsto&\! \tilde C_{\pi_1}\ :=\ g^{-1}\acton C_{\pi_1}-\tilde \nabla_{\pi_1}^0\big(\Sigma_{\pi_1}-\tfrac12\{\Lambda_{\pi_1},\Lambda_{\pi_1}\}\big) +\notag \\
  &&\kern1.5cm+\,\{\tilde B_{\pi_1},\Lambda_{\pi_1}\}+\{\Lambda_{\pi_1},\tilde B_{\pi_1}\}+\{\Lambda_{\pi_1},\tilde\nabla_{\pi_1}\Lambda_{\pi_1}+\tfrac12[\Lambda_{\pi_1},\Lambda_{\pi_1}]\}~,\label{eq:GTC}
\end{eqnarray}
where we have used the abbreviations
\begin{equation}
 \tilde \nabla_{\pi_1}\ :=\ \dd_{\pi_1}+\tilde A_{\pi_1}\acton\eand
  \tilde \nabla_{\pi_1}^0\ :=\ \dd_{\pi_1}+\big(\tilde A_{\pi_1}+\sft(\Lambda_{\pi_1})\big)\acton~.
\end{equation}
\end{subequations}
We shall refer to the transformations \eqref{eq:GTConRelForms} as {\it gauge transformations} of the relative connective structure. We will demonstrate momentarily that these transformations will correspond to certain space-time gauge transformations. The gauge transformations  \eqref{eq:GTConRelForms} then imply that a {\it pure gauge configuration} is one for which the relative connection forms are of the form
\begin{subequations}\label{eq:PureGaugeRelForms}
\begin{eqnarray}
 \kern-.5cm A_{\pi_1}\! &=&\! g^{-1}\dd_{\pi_1}g-\sft(\Lambda_{\pi_1})~,\\
 \kern-.5cm B_{\pi_1}\! &=&\! -\dd_{\pi_1}\Lambda_{\pi_1}-(g^{-1}\dd_{\pi_1}g)\acton\Lambda_{\pi_1}+\tfrac12\sft(\Lambda_{\pi_1})\acton \Lambda_{\pi_1}-\sft(\Sigma_{\pi_1})~,\\
 \kern-.5cm C_{\pi_1}\! &=&\! -\dd_{\pi_1}\big(\Sigma_{\pi_1}+\tfrac12\{\Lambda_{\pi_1},\Lambda_{\pi_1}\}\big)-\notag\\
     &&\kern1cm-\,\big(g^{-1}\dd_{\pi_1}g-\sft(\Lambda_{\pi_1})\big)\acton(\Sigma_{\pi_1}+\tfrac12\{\Lambda_{\pi_1},\Lambda_{\pi_1}\}\big)-\notag\\
      &&\kern1cm-\,\big\{\Lambda_{\pi_1},\dd_{\pi_1}\Lambda_{\pi_1}+\big(g^{-1}\dd_{\pi_1}g-\tfrac12\sft(\Lambda_{\pi_1})\big)\acton\Lambda_{\pi_1}-\tfrac12\langle\Lambda_{\pi_1},\Lambda_{\pi_1}\rangle \big\}~.
\end{eqnarray}
\end{subequations}
This should be compared with the expressions \eqref{eq:DefOf-A}, \eqref{eq:DefOf-B}, and \eqref{eq:DefOf-C} for $A_a$, $B_a$, and $C_a$, respectively, which again justifies calling a relative connective structure relative flat.

Note that the coordinate-patch-dependent $\Theta_a$-transformations appearing in \eqref{eq:GaugeFreedomSmallForms} drop out of \eqref{eq:GTConRelForms} as they should since the relative connection forms are globally defined. At this point we would like to point out that there are additional transformations that leave the splitting \eqref{eq:HolSplitCorr} invariant. They are
\begin{subequations}\label{eq:equiv_of_equivs1}
\begin{eqnarray}
 \kern-1cm g_a\! &\mapsto& \! \tilde g_a\ :=\ g_a\sft(h_a)~,\\
  \kern-1cm  h_{ab}\! &\mapsto& \! \tilde h_{ab}\ :=\ (g_a\acton h_ah_b^{-1})h_{ab}~,\\
  \kern-1cm  \ell_{abc}\! &\mapsto& \!\tilde\ell_{abc}\ :=\ \big[(g_{ab}\acton h_{bc}^{-1})h_{ab}^{-1}(g_a\acton h_c^{-1}h_b)\big]\acton\big\{h_{ab},(\sft(h_{ab}^{-1})g_a)\acton h_bh_c^{-1}\big\}\ell_{abc}
\end{eqnarray}
\end{subequations} 
for some smooth functions $h_a:U_a\to\sH$, and we will come back to them in Section \ref{sec:HGT}. However, also these coordinate-patch-dependent transformations necessarily leave the global relative connection forms $A_{\pi_1}$, $B_{\pi_1}$, and $C_{\pi_1}$ invariant. Thus, the freedom in defining the relative connection forms is given by $g\in H^0(F^{9|8n},\sG)$, $\Lambda_{\pi_1}\in H^0(F^{9|8n},\Omega^1_{\pi_1}\otimes\frh)$, and $\Sigma_{\pi_1}\in H^0(F^{9|8n},\Omega^2_{\pi_1}\otimes\frl)$.

 The induced transformations of the associated curvature forms \eqref{eq:DefOfF},  \eqref{eq:DefOfH}, and  \eqref{eq:DefOfG} read as 
\begin{subequations}
\begin{eqnarray}
 \kern-1cm F_{\pi_1}\! &\mapsto&\! \tilde F_{\pi_1}\ :=\ g^{-1}F_{\pi_1} g-\sft\big(\tilde \nabla_{\pi_1}\Lambda_{\pi_1}+\tfrac12\sft(\Lambda_{\pi_1})\acton \Lambda_{\pi_1}\big)~,\label{eq:GTF}\\
 \kern-1cm  H_{\pi_1}\! &\mapsto&\! \tilde H_{\pi_1}\ :=\ g^{-1}\acton H_{\pi_1}-\big(\tilde F_{\pi_1}-\sft(\tilde B_{\pi_1})\big)\acton\Lambda_{\pi_1}+\notag\\
 &&\kern2cm+\, \sft\Big[-\tilde \nabla_{\pi_1}^0\big(\Sigma_{\pi_1}-\tfrac12\{\Lambda_{\pi_1},\Lambda_{\pi_1}\}\big) +\notag \\
  &&\kern3cm+\,\{\tilde B_{\pi_1},\Lambda_{\pi_1}\}+\{\Lambda_{\pi_1},\tilde B_{\pi_1}\}+\notag\\
  &&\kern3cm +\,\{\Lambda_{\pi_1},\tilde\nabla_{\pi_1}\Lambda_{\pi_1}+\tfrac12[\Lambda_{\pi_1},\Lambda_{\pi_1}]\}\Big]~,\label{eq:GTH}\\
G_{\pi_1}\! &\mapsto&\!  \tilde G_{\pi_1}\ :=\ g^{-1}\acton G_{\pi_1}-\big(\tilde F_{\pi_1}-\sft(\tilde B_{\pi_1})\big)\acton\big(\Sigma_{\pi_1}-\tfrac12\{\Lambda_{\pi_1},\Lambda_{\pi_1}\}\big)+\notag\\
   &&\kern2cm+\,\{\Lambda_{\pi_1},\tilde H_{\pi_1}-\sft(\tilde C_{\pi_1})\}-\{\tilde H_{\pi_1}-\sft(\tilde C_{\pi_1}),\Lambda_{\pi_1}\}-\notag\\
   &&\kern2cm-\,\{\Lambda_{\pi_1},\big(\tilde F_{\pi_1}-\sft(\tilde B_{\pi_1})\big)\acton\Lambda_{\pi_1}\}~.\label{eq:GTG}
\end{eqnarray}
\end{subequations}
The first two equations imply that the fake curvature relations \eqref{eq:DefOfFH} behave covariantly under gauge transformations, that is, $\tilde F_{\pi_1}=\sft(\tilde B_{\pi_1})$ and $\tilde H_{\pi_1}=\sft(\tilde C_{\pi_1})$. In addition, provided that these equations hold, the transformation law of the 4-form curvature $G_{\pi_1}$ simplifies to $G_{\pi_1}\mapsto\tilde G_{\pi_1}=g^{-1}\acton G_{\pi_1}$. This behaviour under gauge transformations explains our definition \eqref{eq:DefOfG} of $G_{\pi_1}$. Note that since $G_{\pi_1}=0$ also $\tilde G_{\pi_1}=0$ confirming again the consistency of our constructions.

\paragraph{Field expansions.}
In the final step of our Penrose--Ward transform, we wish to push down to chiral superspace $M^{6|8n}$ the bundle $E\to F^{9|8n}$ and its relative connective structure. This amounts to `integrating out' the $\PP^3$-dependence in the relative connection forms stemming from the fibres of  $\pi_2:F^{9|8n}\to  M^{6|8n}$.\footnote{Technically speaking, we compute the zeroth direct images of the sheaf $\Omega^r_{\pi_1}$  as in \cite{Saemann:2011nb}.} Eventually, we will obtain a holomorphic principal 3-bundle $E'\to M^{6|8n}$ (which is holomorphically trivial as $M^{6|8n}$ has trivial topology) with a connective structure that is subjected to certain (superspace) constraints. That is, certain components of the associated curvature forms on $E'$ will vanish.

Concretely, the relative connection forms $(A_{\pi_1},B_{\pi_1},C_{\pi_1})$ and the associated curvature forms  $(F_{\pi_1},H_{\pi_1},G_{\pi_1})$ are expanded as
\begin{subequations}\label{eq:RelConStrucExp}
\begin{eqnarray}
 A_{\pi_1}\! &=&\! e_{[A}\lambda_{B]}\,  A^{AB}+e^A_I\, A^I_A~,\\
 B_{\pi_1}\! &=&\! -\tfrac14e_A\wedge e_B\lambda_C\, \varepsilon^{ABCD} B_D{}^{E}\lambda_{E}+\tfrac12 e_A\lambda_B\wedge e^{E}_I\,\varepsilon^{ABCD}\,B_{CD}{}^I_E~+\notag \\
   &&\kern1cm+\,\tfrac12 e^{A}_I\wedge e^{B}_J\, B^{IJ}_{AB}~,\\
  C_{\pi_1}\! &=&\!  -\tfrac13 e_A\wedge e_B\wedge e_C\lambda_D\varepsilon^{ABCD}\,C^{EF}\lambda_E\lambda_F~-\notag\\
 &&\kern1cm -\,\tfrac14 e_A\wedge e_B\lambda_C\, \varepsilon^{ABCD}\wedge e^{E}_I\, (C_{D}{}^F{}^I_E)_0\lambda_F~+\notag\\
 &&\kern1cm +\, \tfrac14 e_A\lambda_B\wedge e^{E}_I\wedge e^{F}_J\,\varepsilon^{ABCD}\,(C_{CD}{}^{IJ}_{EF})_0~+\notag\\
 &&\kern1cm +\, \tfrac16 e^{A}_I\wedge e^{B}_J\wedge e^{C}_K\, C^{IJK}_{ABC}
 \end{eqnarray}
\end{subequations}
and 
\begin{subequations}\label{eq:RelCurvExp}
\begin{eqnarray}
 F_{\pi_1}\! &=&\! -\tfrac14e_A\wedge e_B\lambda_C\, \varepsilon^{ABCD} F_D{}^{E}\lambda_{E}+\tfrac12 e_A\lambda_B\wedge e^{E}_I\,\varepsilon^{ABCD}\,F_{CD}{}^I_E~+\notag \\
   &&\kern1cm+\,\tfrac12 e^{A}_I\wedge e^{B}_J\, F^{IJ}_{AB}~,
 \end{eqnarray}
 \begin{eqnarray}
  H_{\pi_1}\! &=&\!  -\tfrac13 e_A\wedge e_B\wedge e_C\lambda_D\varepsilon^{ABCD}\,H^{EF}\lambda_E\lambda_F~-\notag\\
 &&\kern1cm -\,\tfrac14 e_A\wedge e_B\lambda_C\, \varepsilon^{ABCD}\wedge e^{E}_I\, (H_{D}{}^F{}^I_E)_0\lambda_F~+\notag\\
 &&\kern1cm +\, \tfrac14 e_A\lambda_B\wedge e^{E}_I\wedge e^{F}_J\,\varepsilon^{ABCD}\,(H_{CD}{}^{IJ}_{EF})_0~+\notag\\
 &&\kern1cm +\, \tfrac16 e^{A}_I\wedge e^{B}_J\wedge e^{C}_K\, H^{IJK}_{ABC}~,\\
  G_{\pi_1}\! &=&\!  -\tfrac13 e_A\wedge e_B\wedge e_C\lambda_D\varepsilon^{ABCD}\wedge e^{E}_I\,(G^{FG}{}_E^I)_0\lambda_F\lambda_G~-\notag\\
 &&\kern1cm -\,\tfrac14 e_A\wedge e_B\lambda_C\, \varepsilon^{ABCD}\wedge e^{E}_I\wedge e^{F}_J\, (G_{D}{}^{G}{}^{IJ}_{EF})_0\lambda_{G}~+\notag\\
 &&\kern1cm +\, \tfrac14 e_A\lambda_B\wedge e^{E}_I\wedge e^{F}_J\wedge e^{G}_K\,\varepsilon^{ABCD}\,(G_{CD}{}^{IJK}_{EFG})_0~+\notag\\
 &&\kern1cm +\, \tfrac16 e^{A}_I\wedge e^{B}_J\wedge e^{C}_K\wedge e^{D}_L\, G^{IJKL}_{ABCD}~,
 \end{eqnarray}
\end{subequations}
as  $(A_{\pi_1},B_{\pi_1},C_{\pi_1})$ and  $(F_{\pi_1},H_{\pi_1},G_{\pi_1})$ are of homogeneity zero in the $\lambda_A$ coordinates. Here, we have used the relative differential 1-forms $e_A$ and $e_I^{A}$, which were introduced in \eqref{eq:DefOfRelDer}. In addition, the component $(C_{A}{}^B{}^I_C)_0$ of $C_{\pi_1}$ represents the totally trace-less part of $C_{A}{}^B{}^I_C$ and $(C_{AB}{}^{IJ}_{CD})_0$ denotes the part of $C_{AB}{}^{IJ}_{CD}$ that does not contain terms proportional to $\varepsilon_{ABCD}$. Similar conventions have been used for the components of $H_{\pi_1}$ and  $G_{\pi_1}$.  We would like to emphasise that all $\lambda$-dependence has been made explicit in the above expansions, that is, the component fields in \eqref{eq:RelConStrucExp} and \eqref{eq:RelCurvExp} are superfields defined on the chiral superspace $M^{6|8n}$. 

To clarify the meaning of these component fields, we recall the components of general low-degree differential forms on $M^{6|8n}$. In the following, $(\cdot)$ and $[\cdot]$ denote symmetric and antisymmetric indices, respectively. Note that `fermionic' index pairs $({}^I_A)$ etc.\ always appear totally symmetrised. A 1-form $A$ on $M^{6|8n}$ has the components
\begin{subequations}\label{eq:rFormsOnSupSpa}
\begin{equation}
  \big(A_{[AB]}=\tfrac12\varepsilon_{ABCD}A^{[CD]}~,~A^I_B\big)
\end{equation}
in spinor notation while a  differential 2-form $B$ has the components 
\begin{equation}
  \big(B_A{}^B~,~B_{[AB]}{}^I_C~,~ B^{IJ}_{AB}\big)~,
\end{equation}
where $B_A{}^B$ is trace-less, and a differential 3-form $C$ has the spinor components
\begin{equation}
  \big(C_{(AB)}~,~C^{(AB)}~,~C_{A}{}^B{}^I_C~,~C_{[AB]}{}^{IJ}_{CD}~,~C^{IJK}_{ABC}\big)~,
\end{equation}
where $C_{A}{}^B{}^I_C$ is trace-less over the $AB$ indices. The $C_{(AB)}$-component represents the self-dual part of the purely bosonic components of $C$ while  the $C^{(AB)}$-component represents the anti-self-dual part, respectively. A differential 4-form $D$ has the following spinor components:
\begin{equation}
\begin{aligned}
 \big( D_A{}^B~,~D_{(AB)}{}^I_C~,~D^{(AB)}{}^I_C~,~D_A{}^B{}^{IJ}_{CD}~,~D_{[AB]}{}^{IJK}_{CDE}~,~D_{ABCD}^{IJKL}\big)~,
 \end{aligned}
\end{equation}
\end{subequations}
where $D_A{}^B$ and $D_A{}^B{}^{IJ}_{CD}$ are trace-less over the $AB$ indices.
Hence,  the component fields in \eqref{eq:RelConStrucExp} and \eqref{eq:RelCurvExp}  are nothing but the spinor components of certain differential-form-fields on $M^{6|8n}$. We thus see that the components of the relative connection forms $A_{\pi_1}$ and $B_{\pi_1}$ correspond to connection forms $A$ and $B$ on a holomorphically trivial principal 3-bundle $E'\to M^{6|8n}$. However, as is apparent from \eqref{eq:rFormsOnSupSpa}, we do not obtain all possible components of a connection 3-form $C$  from the expansion  \eqref{eq:RelConStrucExp}. In fact, from $C_{\pi_1}$ we only obtain the components $C^{AB}$, $(C_{A}{}^B{}^I_C)_0$, $(C_{AB}{}^{IJ}_{CD})_0$, and $C^{IJK}_{ABC}$. We shall denote the 3-form on $M^{6|8n}$ that contains only those components by $C_0$.  The `missing' components in the field expansions of the relative curvature forms \eqref{eq:RelCurvExp} will shortly be seen  as part of the constraint equations which the curvature forms $F:=\dd A+\tfrac12[A,A]$, $H:=\dd B+A\acton B$, and $G:=\dd C_0+A\acton C_0+\{B,B\}$ associated with the connective structure $(A,B,C_0)$ have to obey.

\paragraph{Constraint equations on chiral superspace.}
So far, we have obtained a holomorphically trivial principal 3-bundle $E'\to M^{6|8n}$ over chiral superspace with a connective structure that is represented by the connection forms $A$, $B$, and $C_0$ given by the components fields of the expansions \eqref{eq:RelConStrucExp}. Because the relative fake curvatures defined in equation \eqref{eq:DefOfFH} and the relative 4-form curvature $G_{\pi_1}$ vanish, certain components of the associated curvature forms $F$, $H$, and $G$ will vanish. Concretely, the connective structure $(A,B,C_0)$ on $E'\to M^{6|8n}$ is subject to the following set of superspace constraint equations: 
\begin{subequations}\label{eq:constrainteq}
\begin{equation}\label{eq:constrainteqF}
  F_A{}^B\ =\ \sft(B_A{}^B)~,\quad
   F_{AB}{}^I_C\ =\ \sft(B_{AB}{}^I_C)~,\quad
    F^{IJ}_{AB}\ =\ \sft(B^{IJ}_{AB})~,
    \end{equation}
and
 \begin{equation}\label{eq:constrainteqH0}
 \begin{aligned}
  H^{AB}\ =\ \sft(C^{AB})~,\quad
  (H_{A}{}^B{}^I_C)_0\ =\ \sft\big((C_{A}{}^B{}^I_C)_0\big)~,\kern.3cm\\
    (H_{AB}{}^{IJ}_{CD})_0\ =\ \sft\big((C_{AB}{}^{IJ}_{CD})_0\big)~,\quad
      H^{IJK}_{ABC}\ =\ \sft(C^{IJK}_{ABC})~,
      \end{aligned}
      \end{equation}
and
\begin{equation}\label{eq:constrainteqG0}
 (G^{AB}{}^I_C)_0\ =\ 0~,\quad
 (G_A{}^B{}^{IJ}_{CD})_0\ =\ 0~,\quad
 (G_{AB}{}^{IJK}_{CDE})_0\ =\ 0~,\quad
 G^{IJKL}_{ABCD}\ =\ 0~.
\end{equation}
\end{subequations}

The totally `trace-less parts' $(H_{A}{}^B{}^I_C)_0$ and $(H_{AB}{}^{IJ}_{CD})_0$ of  the component of the curvature 3-form $H$ may be written as
\begin{subequations}\label{eq:tracefreepartH}
\begin{eqnarray}
(H_{A}{}^B{}^I_C)_0\! &=&\! H_{A}{}^B{}^I_C-(\delta^B_C\psi^I_A-\tfrac14\delta^B_A\psi^I_C)~,\\
(H_{AB}{}^{IJ}_{CD})_0\! &=&\!  H_{AB}{}^{IJ}_{CD}-\varepsilon_{ABCD}\phi^{IJ}~,
\end{eqnarray}
\end{subequations}
where the fermionic (Gra{\ss}mann-odd) $\psi^I_A$ and bosonic (Gra{\ss}mann-even) $\phi^{IJ}=-\phi^{JI}$ fields represent the `trace-parts'. These fields will turn out later to be the fermions and the scalars of the tensor multiplet. Using these expressions, the constraint equations \eqref{eq:constrainteqH0} thus take the equivalent form
\begin{subequations}\label{eq:constrainteqH}
  \begin{eqnarray}
  H^{AB}\! &=&\! \sft(C^{AB})~,\\
  H_{A}{}^B{}^I_C\! &=&\! \delta^B_C\psi^I_A-\tfrac14\delta^B_A\psi^I_C+\sft\big((C_{A}{}^B{}^I_C)_0\big)~,\\
  H_{AB}{}^{IJ}_{CD}\! &=&\! \varepsilon_{ABCD}\phi^{IJ}+\sft\big((C_{AB}{}^{IJ}_{CD})_0\big)~,\\
  H^{IJK}_{ABC}\! &=&\! \sft(C^{IJK}_{ABC})~.
\end{eqnarray}
\end{subequations}

The components of the curvatures $F$ and $H$ appearing in the constraint equations \eqref{eq:constrainteqF} and \eqref{eq:constrainteqH}  read explicitly as
\begin{subequations}\label{eq:STcurvatures}
\begin{eqnarray}
  F_A{}^B\! &=&\! \partial^{BC} A_{CA}-\partial_{CA}A^{BC}+[A^{BC},A_{CA}]~,\\
  F_{AB}{}^I_C\! &=&\! \partial_{AB}A^I_C-D^I_CA_{AB}+[A_{AB},A^I_C]~,\\
  F^{IJ}_{AB}\! &=&\! D^I_AA^J_B+D^J_B A^I_A+[A^I_A,A^J_B]+4\Omega^{IJ}A_{AB}~,\\
  H^{AB}\! &=&\! \nabla^{C(A}B_C{}^{B)}~,\\
  H_{A}{}^B{}^I_C\! &=&\!\nabla^I_CB_A{}^B-\nabla^{DB}B_{DA}{}^I_C+\nabla_{DA}B^{DB}{}^I_C~,\\
  H_{AB}{}^{IJ}_{CD}\! &=&\!\nabla_{AB}B^{IJ}_{CD}-\nabla^I_C B_{AB}{}^J_D-\nabla^J_D B_{AB}{}^I_C-\notag\\
    &&\kern1cm-\,2\Omega^{IJ}(\varepsilon_{ABF[C} B_{D]}{}^F-\varepsilon_{CDF[A} B_{B]}{}^F)~,\\
  H^{IJK}_{ABC}\! &=&\! \nabla^I_A B^{JK}_{BC}+\nabla^J_BB^{IK}_{AC}+\nabla^K_C B^{IJ}_{AB}+\notag\\
     &&\kern1cm+\,4\Omega^{IJ}B_{AB}{}^K_C+4\Omega^{IK}B_{AC}{}^J_B+4\Omega^{JK}B_{BC}{}^I_A~.
\end{eqnarray}
\end{subequations}
These equations follow from \eqref{eq:DefOfFH} and the expansions \eqref{eq:RelConStrucExp} and \eqref{eq:RelCurvExp} together with the relations \eqref{eq:MCSE}; these components also follow directly from the expressions $F=\dd A+\tfrac12[A,A]$ and $H=\dd B+A\acton B$ on chiral superspace. The self-dual part $H_{AB}$ of the 3-form curvature is then given by 
\begin{equation}\label{eq:DefOfSDH}
   H_{AB}\ :=\ \nabla_{C(A}B_{B)}{}^C~.
\end{equation}

The `trace-less parts' $(G^{AB}{}^I_C)_0$, $(G_A{}^B{}^{IJ}_{CD})_0$, and $(G^{AB}{}^{IJK}_{CDE})_0$ of the components of the curvature 4-form $G$ may be written as
\begin{subequations}
\begin{eqnarray}
 \kern-.5cm (G^{AB}{}^I_C)_0\! &=&\! ~G^{AB}{}^I_C-\chi^{I(A}\delta^{B)}_C~,\\
 \kern-.5cm (G_A{}^B{}^{IJ}_{CD})_0\! &=&\! G_A{}^B{}^{IJ}_{CD}-\big(U^{IJ}_{A[C}\delta^B_{D]}+\tfrac14\delta^A_BU^{IJ}_{[CD]}\big)-\big(V^{IJ}_{A(C}\delta^B_{D)}-\tfrac14\delta^A_BV^{IJ}_{(CD)}\big)~,\\
\kern-.5cm (G_{AB}{}^{IJK}_{CDE})_0\! &=&\! G_{AB}{}^{IJK}_{CDE}-\varepsilon_{ABCD}\tilde\psi_{E}^{IJK}-\notag\\
&&~~~~-\,5\mbox{ terms to totally symmetrise in }({}^I_A)({}^J_B)({}^K_C)~,
\end{eqnarray}
\end{subequations}
where $\chi^{IA}$ and $\tilde\psi_{A}^{IJK}=\tilde\psi_{A}^{[IJ]K}$ are fermionic (Gra{\ss}mann-parity odd) and $U^{IJ}_{AB}=U^{[IJ]}_{AB}$ and $V^{IJ}_{AB}=V^{(IJ)}_{AB}$ are bosonic (Gra{\ss}mann-parity even) which represent the `trace-parts'. The bi-spinor $U^{IJ}_{AB}$ can be decomposed into a vector $U^{IJ}_{[AB]}$ and a self-dual 3-form $U^{IJ}_{(AB)}$ and similarly for $V^{IJ}_{AB}$. Thus, the constraint equations \eqref{eq:constrainteqG0} read as
\begin{equation}\label{eq:constrainteqG}
\begin{aligned}
G^{AB}{}^I_C\ =\ \chi^{I(A}\delta^{B)}_C~,\quad
G_{AB}{}^{IJK}_{CDE}\ =\ \varepsilon_{ABCD}\tilde\psi_{E}^{IJK}+\mbox{symmetrisation}~,\kern.8cm\\
 G^{IJKL}_{ABCD}\ =\ 0~,\quad
G_A{}^B{}^{IJ}_{CD}\ =\ \big(U^{IJ}_{A[C}\delta^B_{D]}+\tfrac14\delta^A_BU^{IJ}_{[CD]}\big)+\big(V^{IJ}_{A(C}\delta^B_{D)}-\tfrac14\delta^A_BV^{IJ}_{(CD)}\big)
\end{aligned}
\end{equation}
with the curvature components given by\footnote{Recall that $G=\dd C_0+A\acton C_0+\{B,B\}$.}
\begin{subequations}
\begin{eqnarray}
 G^{AB}{}^I_C\! &=&\! -\nabla^I_C C^{AB}+\tfrac14\nabla^{D(A}(C_D{}^{B)}{}^I_C)_0-\notag\\
 && \kern1cm-\, \tfrac14\{B_D{}^{(A},B^{B)D}{}^I_C\}+\tfrac14\{B^{D(A}{}^I_C,B_D{}^{B)}\}~,\\
 G_A{}^B{}^{IJ}_{CD} \! &=&\! \tfrac12\Big[\Big(\nabla^I_C(C_A{}^B{}^J_D)_0-8\Omega^{IJ}\varepsilon_{ACDE}C^{EB}-\notag\\
  &&\kern1cm-\,\{B^{EB}{}^I_C,B_{EA}{}^J_D\}+\{B_{EA}{}^J_D,B^{EB}{}^I_C\} +({}^I_C)\leftrightarrow({}^J_D)\Big)-\notag\\
  &&\kern1cm-\,\nabla^{EB}(C_{EA}{}^{IJ}_{CD})_0+\nabla_{EA}(C^{EB}{}^{IJ}_{CD})_0+\notag\\
  &&\kern1cm+\,\{B_A{}^B,B^{IJ}_{CD}\}+\{B^{IJ}_{CD},B_A{}^B\}\Big]~,\\
  G^{AB}{}^{IJK}_{CDE}\! &=&\! \tfrac13\Big[\nabla^{AB}C^{IJK}_{CDE}+\notag\\
  &&\kern1cm+\,\Big(-\nabla^I_C(C^{AB}{}^{JK}_{DE})_0+8\Omega^{IJ}\delta_{[C}^{[A}(C_{D]}{}^{B]}{}^K_E)_0+\notag\\
  &&\kern2cm+\,\{B^{AB}{}^I_C,B^{JK}_{DE}\}+\{B^{JK}_{DE},B^{AB}{}^I_C\}+\notag\\
  &&\kern2cm+\,({}^I_C)\leftrightarrow({}^J_D)+({}^I_C)\leftrightarrow({}^K_E)\Big)\Big]~,
  \\
  G^{IJKL}_{ABCD}\! &=&\! \tfrac{1}{4!}\Big[ -\nabla^I_A C^{JKL}_{BCD}+\tfrac32\Omega^{IJ}(C_{AB}{}^{KL}_{CD})_0+\tfrac32\{B^{IJ}_{AB},B_{CD}^{KL}\}+\notag\\
  &&\kern1cm+\,23\mbox{ terms to totally symmetrise in }({}^I_A)({}^J_B)({}^K_C)({}^L_D)\Big]~.
  \end{eqnarray}\end{subequations}
In deriving these equations, we have again made use of \eqref{eq:MCSE}. Note that one can show that all these components of $G$ lie in the kernel of $\sft$.

\paragraph{Gauge freedom on chiral superspace.}
Because of \eqref{eq:GTConRelForms}, there is a gauge freedom in the above constraint equations. In particular, the gauge parameters $\Lambda_{\pi_1}$ and $\Sigma_{\pi_1}$ appearing in the gauge transformations \eqref{eq:GTConRelForms} of the relative connective structure are expanded as
\begin{subequations}\label{eq:GTExp}
\begin{eqnarray}
 \Lambda_{\pi_1}\! &=&\! e_{[A}\lambda_{B]}\,  \Lambda^{AB}+e^{AB}_I\lambda_A\, \Lambda^I_B~,\\
 \Sigma_{\pi_1}\! &=&\! -\tfrac14e_A\wedge e_B\lambda_C\, \varepsilon^{ABCD} \Sigma_D{}^{E}\lambda_{E}+\tfrac12 e_A\lambda_B\wedge e^{EF}_I\lambda_E\,\varepsilon^{ABCD}\,\Sigma_{CD}{}^I_F~+\notag \\
   &&\kern1cm+\,\tfrac12 e^{CA}_I\lambda_C\wedge e^{DB}_J\lambda_D\, \Sigma^{IJ}_{AB}~,
\end{eqnarray}
\end{subequations}
where $ \Sigma_A{}^{B}$ is trace-less.  Note that $g$ in \eqref{eq:GTConRelForms}  is a globally defined, holomorphic  $\sG$-valued function on the correspondence space, and, as such, it does not depend on the coordinates $\lambda_A$ (since $\PP^3$ is compact). Thus, $g$ descents down to $M^{6|8n}$ directly. 

The coefficient functions of  $\Lambda_{\pi_1}$ and $\Sigma_{\pi_1}$ together with $g$ are the gauge parameters on $M^{6|8n}$: upon substituting the expansions \eqref{eq:GTExp} and \eqref{eq:RelConStrucExp} into the transformations \eqref{eq:GTConRelForms}, we find the following gauge transformation on chiral superspace $M^{6|8n}$:
\begin{subequations}\label{eq:GTSTConForms}
\begin{eqnarray}
 \kern-.7cm A_{AB}\! &\mapsto&\! \tilde A_{AB}\ :=\ g^{-1}A_{AB} g+g^{-1}\partial_{AB}g-\sft(\Lambda_{AB})~,\label{eq:GSTTA1}\\
  \kern-.7cm A^I_A\! &\mapsto&\! \tilde A^I_A\ :=\ g^{-1}A^I_A g+g^{-1} D^I_Ag-\sft(\Lambda^I_A)~,\label{eq:GSTTA2}\\[2mm]
 \kern-.7cm  B_A{}^B\! &\mapsto&\! \tilde B_A{}^B\ :=\ g^{-1}\acton B_A{}^B-\tilde \nabla^{BC}\Lambda_{CA}+\tilde \nabla_{CA}\Lambda^{BC}-\notag\\
   && \kern2cm -\, \tfrac12\sft(\Lambda^{BC})\acton \Lambda_{CA}+\tfrac12\sft(\Lambda_{CA})\acton \Lambda^{BC}-\sft(\Sigma_A{}^B)~,\label{eq:GTSTB1}\\
  \kern-.7cm  B_{AB}{}^I_C\! &\mapsto&\! \tilde B_{AB}{}^I_C\ :=\ g^{-1}\acton B_{AB}{}^I_C-\tilde \nabla_{AB}\Lambda_C^I+\tilde\nabla^I_C\Lambda_{AB}-\notag\\
   &&\kern2cm -\, \tfrac12\sft(\Lambda_{AB})\acton \Lambda_C^I+
   \tfrac12\sft(\Lambda_C^I)\acton \Lambda_{AB} -\sft(\Sigma_{AB}{}^I_C)~,\label{eq:GTSTB2}\\
   \kern-.7cm  B^{IJ}_{AB}\! &\mapsto&\! \tilde B^{IJ}_{AB}\ :=\ g^{-1}\acton B^{IJ}_{AB}-\tilde \nabla^I_A\Lambda^J_B-\tilde\nabla^J_B\Lambda^I_A-4\Omega^{IJ}\Lambda_{AB}-\notag\\
   &&\kern2cm-\,\tfrac12\sft(\Lambda^I_A)\acton \Lambda^J_B-\tfrac12\sft(\Lambda^J_B)\acton \Lambda^I_A-\sft(\Sigma^{IJ}_{AB})~,\label{eq:GTSTB3}\\
\kern-.7cm C^{AB}\! &\mapsto&\! \tilde C^{AB}\ :=\ g^{-1}\acton C^{AB}-\tilde\nabla^{0\,C(A}\big(\Sigma_{C}{}^{B)}-\tfrac12\{\Lambda^{B)D},\Lambda_{DC}\}+\tfrac12\{\Lambda_{DC},\Lambda^{B)D}\}\big) -\notag \\
  &&\kern2cm-\,\{\tilde B_{C}{}^{(A},\Lambda^{B)C}\}
   +\{\Lambda^{C(A},\tilde B_{C}{}^{B)}\}+\notag\\
  &&\kern2cm +\, \{\Lambda^{C(A},\tilde \nabla^{B)D}\Lambda_{DC}-\tilde \nabla_{DC}\Lambda^{B)D}+[\Lambda^{B)D},\Lambda_{DC}]\}~,\label{eq:GTSTC1}\\
  \kern-.7cm (C_{A}{}^B{}^I_C)_0\! &\mapsto&\! (\tilde C_{A}{}^B{}^I_C)_0\ :=\ g^{-1}\acton (C_{A}{}^B{}^I_C)_0-\notag\\
   &&\kern2cm-\,\Big[\tilde \nabla_{C}^{0\,I}\big(\Sigma_{A}{}^B-\tfrac12\{\Lambda^{BC},\Lambda_{CA}\}+\tfrac12\{\Lambda_{CA},\Lambda^{BC}\}\big) +\notag \\
  &&\kern2.5cm +\,\tilde \nabla^{0\,DB}\big(\Sigma_{DA}{}^I_C-\tfrac12\{\Lambda_{DA},\Lambda^{I}_C\}+\tfrac12\{\Lambda^I_C,\Lambda_{DA}\}\big)+\notag\\
   &&\kern2.5cm +\,\tilde \nabla^{0}_{DA}\big(\Sigma^{DB}{}^I_C-\tfrac12\{\Lambda^{DB},\Lambda^{I}_C\}+\tfrac12\{\Lambda^I_C,\Lambda^{DB}\}\big)-\notag\\
  &&\kern2.5cm-\,\{\tilde B_{A}{}^B,\Lambda^I_C\}-\{\Lambda^I_C,\tilde B_{A}{}^B\}+\{\tilde B_{DA}{}^I_C,\Lambda^{DB}\}-\{\Lambda^{DB},\tilde B_{DA}{}^I_C\}+\notag\\
    &&\kern2.5cm+\,\{\tilde B^{DB}{}^I_C,\Lambda_{DA}\}-\{\Lambda_{DA},\tilde B^{DB}{}^I_C\}-\notag\\
  &&\kern2.5cm-\, \{\Lambda^I_C,\tilde \nabla^{BD}\Lambda_{DA}-\tilde \nabla_{DA}\Lambda^{BD}+[\Lambda^{BD},\Lambda_{DA}]\}\Big]_0~,\label{eq:GTSTC2}\\
  \kern-.7cm (C_{AB}{}^{IJ}_{CD})_0\! &\mapsto&\! (\tilde C_{AB}{}^{IJ}_{CD})_0\ :=\ g^{-1}\acton (C_{AB}{}^{IJ}_{CD})_0-\notag\\
   &&\kern2cm-\,\Big[ \tilde \nabla_{AB}^0\big(\Sigma_{CD}^{IJ}-\tfrac12\{\Lambda_{C}^I,\Lambda_{D}^J\}-\tfrac12\{\Lambda_{D}^J,\Lambda_{C}^I\}\big) -\notag \\
  &&\kern2.5cm -\,\tilde \nabla_{C}^{0\,I}\big(\Sigma_{AB}{}^J_D-\tfrac12\{\Lambda_{AB},\Lambda_{D}^J\}+\tfrac12\{\Lambda_{D}^J,\Lambda_{AB}\}\big) -\notag \\
   &&\kern2.5cm -\,\tilde \nabla_{D}^{0\,J}\big(\Sigma_{AB}{}^I_C-\tfrac12\{\Lambda_{AB},\Lambda_{C}^I\}+\tfrac12\{\Lambda_{C}^I,\Lambda_{AB}\}\big) -\notag \\
    &&\kern2.5cm -\, 2\Omega^{IJ}\varepsilon_{ABF[C} \big(\Sigma_{D]}{}^F+\tfrac12\{\Lambda^{FG},\Lambda_{B]G}\}-\tfrac12\{\Lambda_{D]G},\Lambda^{AG}\}\big)+\notag\\
   &&\kern2.5cm +\, 2\Omega^{IJ}\varepsilon_{CDF[A} \big(\Sigma_{B]}{}^F+\tfrac12\{\Lambda^{FG},\Lambda_{B]G}\}-\tfrac12\{\Lambda_{B]G},\Lambda^{AG}\}\big)-\notag\\
  &&\kern2.5cm-\,\{\tilde B_{CD}^{IJ},\Lambda_{AB}\}-\{\Lambda_{AB},\tilde B_{CD}^{IJ}\}-\{\tilde B_{AB}{}^J_D,\Lambda_{C}^I\}+\{\Lambda_{C}^I,\tilde B_{AB}{}^J_D\}-\notag\\
    &&\kern2.5cm-\,\{\tilde B_{AB}{}^I_C,\Lambda_{D}^J\}+\{\Lambda_{D}^J,\tilde B_{AB}{}^I_C\}-\notag\\
    &&\kern2.5cm-\,\{\Lambda_{AB},\tilde \nabla^I_C\Lambda^J_D+\tilde\nabla^J_D\Lambda^I_C-4\Omega^{IJ}\Lambda_{CD}+[\Lambda_C^I,\Lambda_{D}^J]\}+\notag\\
     &&\kern2.5cm+\,\{\Lambda_{C}^I,\tilde \nabla_{AB}\Lambda_D^J-\tilde\nabla^J_D\Lambda_{AB}+[\Lambda_{AB},\Lambda_{C}^I]\}+\notag\\
 &&\kern2.5cm+\,\{\Lambda_{D}^J,\tilde \nabla_{AB}\Lambda_C^I-\tilde\nabla^I_C\Lambda_{AB}+[\Lambda_{AB},\Lambda_D^J]\}\Big]_0~,\label{eq:GTSTC3}
 \end{eqnarray}

\begin{eqnarray}   
  \kern-.7cm C^{IJK}_{ABC}\! &\mapsto&\! \tilde C^{IJK}_{ABC}\ :=\ g^{-1}\acton C^{IJK}_{ABC}+\notag\\
  &&\kern2cm -\,\tilde \nabla_{A}^{0\,I}\big(\Sigma_{BC}^{JK}-\tfrac12\{\Lambda_{B}^J,\Lambda_{C}^K\}-\tfrac12\{\Lambda_{C}^K,\Lambda_{B}^J\}\big) +\notag \\
   &&\kern2cm -\,\tilde \nabla_{B}^{0\,J}\big(\Sigma_{AC}^{IK}-\tfrac12\{\Lambda_{A}^I,\Lambda_{C}^K\}-\tfrac12\{\Lambda_{C}^K,\Lambda_{A}^I\}\big) +\notag \\
    &&\kern2cm -\,\tilde \nabla_{C}^{0\,K}\big(\Sigma_{AB}^{IJ}-\tfrac12\{\Lambda_{A}^I,\Lambda_{B}^J\}-\tfrac12\{\Lambda_{B}^J,\Lambda_{A}^I\}\big) +\notag \\
    &&\kern2cm -\, 4\Omega^{IJ}\big(\Sigma_{AB}{}^K_C -\tfrac12\{\Lambda_{AB},\Lambda_{C}^K\}+\tfrac12\{\Lambda_{C}^K,\Lambda_{AB}\}\big)+\notag\\
    &&\kern2cm -\, 4\Omega^{IK}\big(\Sigma_{AC}{}^J_B -\tfrac12\{\Lambda_{AC},\Lambda_{J}^B\}+\tfrac12\{\Lambda_{B}^J,\Lambda_{AC}\}\big)+\notag\\
    &&\kern2cm -\, 4\Omega^{JK}\big(\Sigma_{BC}{}^I_A -\tfrac12\{\Lambda_{BC},\Lambda_{A}^I\}+\tfrac12\{\Lambda_{A}^I,\Lambda_{BC}\}\big)+\notag\\
 &&\kern2cm+\,\{\tilde B_{AB}^{IJ},\Lambda_C^K\}+\{\Lambda_C^K,\tilde B_{AB}^{IJ}\}-\notag\\
 &&\kern2cm+\,\{\tilde B_{AC}^{IK},\Lambda_B^J\}+\{\Lambda_B^J,\tilde B_{AC}^{IK}\}-\notag\\
 &&\kern2cm+\,\{\tilde B_{BC}^{JK},\Lambda_A^I\}+\{\Lambda_A^I,\tilde B_{BC}^{JK}\}-\notag\\
  &&\kern2cm-\,\{\Lambda_A^I,\tilde \nabla^J_B\Lambda^K_C+\tilde\nabla^K_C\Lambda^J_B+4\Omega^{JK}\Lambda_{BC}+[\Lambda^J_B,\Lambda^K_C]\}-\notag\\
  &&\kern2cm-\,\{\Lambda_B^J,\tilde \nabla^I_A\Lambda^K_C+\tilde\nabla^K_C\Lambda^I_A+4\Omega^{IK}\Lambda_{AC}+[\Lambda^I_A,\Lambda^K_C]\}-\notag\\
  &&\kern2cm-\,\{\Lambda_C^K,\tilde\nabla^I_A\Lambda^J_B+\tilde \nabla^J_B\Lambda^I_A+4\Omega^{IJ}\Lambda_{AB}+[\Lambda^I_A,\Lambda^J_B]\}~.\label{eq:GTST4}
\end{eqnarray}
\end{subequations}

\subsection{Discussion of the constraint equations}\label{sec:discussion}

To summarise the discussion of the previous section, by starting from an $M^{6|8n}$-trivial holomorphic principal 3-bundle $\hat E$ over the twistor space $P^{6|2n}$, we have constructed a holomorphically trivial principal 3-bundle over chiral superspace $M^{6|8n}$ that comes equipped with a holomorphic connective structure subjected to the superspace constraint equations \eqref{eq:constrainteqF}, \eqref{eq:constrainteqH}, and \eqref{eq:constrainteqG}. In particular, the \v Cech equivalence class of any such bundle over the twistor space gives a gauge equivalence class of complex holomorphic solutions to these constraint equations. The inverse of this Penrose--Ward transform is well-defined, and returns an $M^{6|8n}$-trivial holomorphic principal 3-bundle $\hat E'$ over the twistor space $P^{6|2n}$ that is equivalent to $\hat E$. To see this, we take the components of the connective structure on $M^{6|8n}$ and construct the relative connective structure using equations \eqref{eq:RelConStrucExp}. The fact that the relative 4-form curvature  as well as the relative fake curvatures vanish, implies that the relative connective structure is pure gauge. From this observation, the reverse construction of the \v{C}ech cocycles describing the principal 3-bundle $\hat E'$ over twistor space is essentially straightforward. We may therefore formulate the following theorem:
\pagebreak
{\theorem
There is a bijection between

\begin{center}
\begin{minipage}{14cm}
\begin{itemize}
 \setlength{\itemsep}{-1mm}
\item[{\rm (i)}] equivalence classes of $M^{6|8n}$-trivial holomorphic principal 3-bundles over the twistor space $P^{6|2n}$ and
\item[{\rm (ii)}] gauge equivalence classes of (complex holomorphic) solutions to the constraint equations \eqref{eq:constrainteq} on chiral superspace  $M^{6|2n}$.
\end{itemize}
\end{minipage}
\end{center}
}

Let us now discuss the constraint equations \eqref{eq:constrainteq}, or, equivalently, \eqref{eq:constrainteqF}, \eqref{eq:constrainteqH}, and \eqref{eq:constrainteqG} in more detail. The equation $H^{AB}=\sft(C^{AB})$ appearing in \eqref{eq:constrainteqH} fixes the anti-self-dual part of $H^{AB}$.\footnote{This is quite similar to what happens in the field equations of \cite{Samtleben:2011fj}.} However, by inspecting  \eqref{eq:GTSTC1}, we realise that we have, in fact, enough gauge freedom to choose a gauge via the gauge transformation \eqref{eq:GTSTC1} in which $C^{AB}$ vanishes identically. Alternatively, this also follows from an analogous cohomological discussion to that presented in the Abelian case in \cite{0198535651,Saemann:2011nb}. This gauge makes then transparent that our constraint equations indeed contain a non-Abelian generalisation of the self-dual tensor multiplet represented by $(H_{AB},\psi^I_A,\phi^{IJ})$, where $H_{AB}$ was defined in \eqref{eq:DefOfSDH} and the spinors $\psi^I_A$ and scalars $\phi^{IJ}$ were given in \eqref{eq:tracefreepartH}.  

Note that, by construction, the fields $(H_{AB},\psi^I_A,\phi^{IJ})$ take values  in the kernel of $\sft:\frh\to\frg$. This is analogous to what was obtained previously in the context of principal 2-bundles \cite{Saemann:2011nb}. However, it is important to realise that contrary to the principal 2-bundle case, here this does {\it not} imply that $(H_{AB},\psi^I_A,\phi^{IJ})$ have to take values in the centre of the Lie algebra $\frh$. Specifically, if, say, $Y_1\in\ker(\sft:\frh\to\frg)$, then by virtue of axiom (ii) of a differential Lie 2-crossed module, we obtain $[Y_1,Y_2]=\langle Y_1,Y_2\rangle\neq 0$, in general, for any $Y_2\in\frh$. Thus, as a result of the non-triviality of the Peiffer lifting, the tensor multiplet $(H_{AB},\psi^I_A,\phi^{IJ})$ obtained from principal 3-bundles is generally non-Abelian.  Furthermore, as our equations are formulated on superspace, they are manifestly supersymmetric. In addition, the whole twistor construction is superconformal. Altogether, we have therefore obtained $\CN=(n,0)$ manifestly superconformal, interacting field theories with $n=0,1,2$ that contain a non-Abelian generalisation of the $\CN=(n,0)$ tensor multiplet. 

We should note that a general gauge theory on a principal 3-bundle over chiral superspace $M^{6|8n}$, which we will discuss in Section \ref{sec:HGT}, will contain the full fake curvature condition $\CH=H-\sft(C)=0$. According to our constraint equations \eqref{eq:constrainteqH}, we do not find the full fake curvature equation on chiral superspace. However, if, say, the sequence $\frl\to\frh\to\frg$ was exact at $\frh$, we could always adjust the 3-form potential $C$ such that the general fake curvature condition holds since, by construction, we have $\sft(H)=0$ for the full 3-form curvature. Otherwise, even though the relation of our constraint equations to parallel transport of two-dimensional objects remains unclear at this stage\footnote{Note that parallel transport requires the vanishing of fake curvatures, see \cite{Girelli:2003ev,Baez:2004in,Baez:0511710,Martins:2009aa}.}, they still describe a consistent superconformal gauge theory.\footnote{In \cite{Martins:2010ry} a (non-supersymmetric) higher gauge theory including an action principle based on principal 2-bundles was obtain that does not require the vanishing of the fake curvature either.}

A particularly interesting point is now the coupling of the `matter fields', such as $\psi^I_A$ and $\phi^{IJ}$ defined in \eqref{eq:tracefreepartH}, to the connective structure. Equations \eqref{eq:constrainteqH} together with \eqref{eq:GSTTA1}--\eqref{eq:GTSTB3} show that these fields transform under gauge transformations as
\begin{equation}
 \psi^I_A\ \mapsto\ \tilde\psi^I_A\ :=\ g^{-1}\acton \psi^I_A+\sft(\alpha^I_A)\eand \phi^{IJ}\ \mapsto\ \tilde\phi^{IJ}\ :=\ g^{-1}\acton\phi^{IJ}+\sft(\alpha^{IJ})~,
\end{equation}
where the gauge parameters $\alpha^I_A$ and $\alpha^{IJ}$ are fixed by the gauge parameters $\Lambda$ and $\Sigma$ entering  \eqref{eq:GSTTA1}--\eqref{eq:GTSTB3}. This is the desired transformation and such a coupling of matter fields to a connective structure of higher gauge theory had only been obtained in \cite{Saemann:2012uq} so far. Notice that $H_{AB}$ transforms likewise as $H_{AB}\mapsto\ \tilde H_{AB}:=g^{-1}\acton H_{AB}+\sft(\alpha_{AB})$.

Finally, let us come to a few special cases of our construction. First of all, if we reduce our principal 3-bundle to a principal 2-bundle by choosing a Lie 2-crossed module $\{\unit\}\rightarrow \sH\rightarrow \sG$, our equations reduce to those obtained in \cite{Saemann:2012uq}. A further reduction to the Abelian case $\{\unit\}\rightarrow \sU(1)\rightarrow\{\unit\}$ then obviously leads to the situation described in \cite{Saemann:2011nb,Mason:2011nw}.  One can also perform the Penrose--Ward transform for principal 3-bundles over the hyperplane twistor space introduced in \cite{Saemann:2011nb}. This will yield solutions to non-Abelian generalisations of the self-dual string equations. As the discussion is straightforward (cf.\ the discussion for principal 2-bundles in \cite{Saemann:2012uq}), we refrain from going into any further details.

\subsection{Constraint equations and superconformal field equations}

A detailed analysis of our constraint equations requires to reduce them to an equivalent set of field equations on six-dimensional space-time $M^6$. This step is well understood e.g.\ for the constraint equations of maximally supersymmetric Yang--Mills theories \cite{Harnad:1984vk,Harnad:1985bc} and three-dimensional supersymmetric Chern--Simons theories \cite{Samtleben:2009ts,Samtleben:2010eu}. 

In this reduction, the components of the curvatures along fermionic directions are identified with matter superfields as done above, and the Bianchi identities yield the corresponding field equations for a supermultiplet of superfields. These field equations can then be shown to be equivalent to the field equations restricted to the purely bosonic part of the superfields. One thus arrives at a set of supersymmetric field equations on ordinary space-time. This reduction procedure, however, is very involved, and it is therefore beyond the scope of this paper and postponed to future work.

Although we do not have the explicit field equations on space-time, we already know that they will consist of a $\CN=(2,0)$ superconformal higher gauge theory involving a connective structure on a trivial principal 3-bundle over six-dimensional Minkowski space. Recall that recently, $\CN=(1,0)$ superconformal field theories were derived from a non-Abelian generalisation of the tensor hierarchy \cite{Samtleben:2011fj,Samtleben:2012mi,Samtleben:2012fb}. An obvious question is now if there is any relation between our $\CN=(2,0)$ field equations and those of the $(1,0)$-models. 

The general relation between higher gauge theories and the $(1,0)$-models was explored in \cite{Palmer:2013pka}. There it was found that the algebraic structure underlying the $(1,0)$-models can be encoded in a certain class of semistrict Lie $n$-algebras. Moreover, the field equations are those of a higher gauge theory with an additional six-dimensional vector multiplet coupled to the tensor multiplet. In \cite{Samtleben:2011fj,Samtleben:2012mi,Samtleben:2012fb}, however, no fake curvature conditions were imposed and correspondingly, there is no underlying parallel transport of extended objects. A further difference to our superconformal field equations is the fact that we had to restrict our discussion to the case of strict Lie $n$-algebras, as a more general cocycle description of principal 3-bundles is not yet available. We thus see that there is a large overlap in field content and equations of motion between our superconformal field equations and the $(1,0)$-models. 

Finally, it is certainly tempting to speculate about the role of our constraint equations in the description of systems of multiple M5-branes. Recall that the interactions of M5-branes are mediated by M2-branes suspended between these, and their boundaries form so-called self-dual strings. An effective description of M5-branes should therefore capture the parallel transport of these self-dual strings, just as the effective description of D-branes by Yang--Mills theories captures the parallel transport of the endpoints of strings. This argument directly leads to superconformal higher gauge theories such as the one constructed in this paper. 

Possibly the most important consistency check for an effective description of M5-branes is a convincing reduction mechanism to five-dimensional supersymmetric Yang--Mills theory. This remains an open problem for both our field equations as well as those of the $(1,0)$-model. Similarly unclear is what higher gauge group one should choose for a description of systems of multiple M5-branes. There are arguments which are based on a relation of higher principal bundles with principal bundles on loop space and the necessity of a reduction to five-dimensional Yang--Mills theory. Naively, they suggest to use the automorphism 2-group of $\sU(N)$. More sophisticated arguments suggest to use string 2-groups, cf.\ \cite{Fiorenza:2012tb}.

\section{Higher gauge theory on principal 3-bundles}\label{sec:HGT}

In the derivation of the constraint equations in the last section, all features of gauge theory on principal 3-bundles have become apparent. Let us summarise these in the following.

To give a complete description, we will discuss the underlying non-Abelian differential cohomology right from the start. Recall that Abelian $(n-1)$-gerbes with connective structure are described by Deligne cohomology \cite{0817647309}. The non-Abelian differential cohomology we have in mind here is one that is based on principal $n$-bundles with structure $n$-groups, and which reduces to Deligne cohomology for the case of Abelian $(n-1)$-gerbes, that is principal $n$-bundles with gauge $n$-group $\sB^{n-1}\sU(1)$. In the following, we shall iterate through the principal $n$-bundles for $n=0,1,2,3$. 

Consider a manifold $M$ with a cover $\frU:=\{U_a\}$. We denote \v{C}ech $p$-cochains that take values in the sheaf of smooth functions into $\sG$ by $C^{p,0}(\frU,\sG)$ and \v{C}ech $p$-cochains that take values in the sheaf of differential $q$-forms on $M$ times a Lie algebra $\frg$ by $C^{p,q}(\frU,\frg)$.

\paragraph{Degree 0.} A degree-0 cochain $\{g_a\}$ with values in $\sG$ is specified by a set $\{g_a\}\in C^{0,0}(\frU,\sG)$. The cocycle condition reads as
\begin{equation}
 g_a\ =\ g_b~~~\mbox{on}~~~U_a\cap U_b~.
\end{equation}
As usual in degree 0, there is no equivalence between cocycles in terms of coboundaries. Thus, an element of the degree 0 cohomology set defines a smooth function $M\rightarrow \sG$, which could be called a {\em principal 0-bundle}.

\paragraph{Degree 1.} A degree-1 $\sG$-valued cochain $(\{g_{ab}\},\{A_a\})$  is given by the \v{C}ech cochains $\{g_{ab}\}\in C^{1,0}(\frU,\sG)$ and $\{A_a\}\in C^{0,1}(\frU,\frg)$. The degree-1 cocycle condition amounts to 
\begin{subequations}
\begin{eqnarray}
 g_{ac}\! &=&\! g_{ab}g_{bc}~~~\mbox{on}~~~U_a\cap U_b\cap U_c~,\\
 A_b\! &=&\! g_{ab}^{-1}A_a g_{ab}+g_{ab}^{-1}\dd g_{ab}~~~\mbox{on}~~~U_a\cap U_b~.
\end{eqnarray}
\end{subequations}
Two degree-1 cocycles $(\{g_{ab}\},\{A_a\})$ and $(\{\tilde{g}_{ab}\},\{\tilde{A}_a\})$ are considered equivalent if there is a degree-0 cochain $\{g_a\}$ with values in $\sG$ such that
\begin{subequations}\label{eq:gauge-trafo-1}
\begin{eqnarray}
 \tilde{A}_a\! &=&\! g_a^{-1}A_ag_a+g_a^{-1}\dd g_a~~~\mbox{on}~~~U_a~,\\
 \tilde{g}_{ab}\! &=&\! g_a^{-1}g_{ab} g_b~~~\mbox{on}~~~U_a\cap U_b~.
\end{eqnarray}
\end{subequations}
We conclude that elements of the degree-1 cohomology set define principal (1-)bundles with connection. Note that the second cocycle condition turns the local Lie algebra valued one-forms $\{A_a\}$ into a global object, the connection. The curvature of $(\{g_{ab}\},\{A_a\})$ ,
\begin{equation}
 F_a\ :=\ \dd A_a+\tfrac{1}{2}[A_a,A_a]~,
\end{equation}
fulfils the Bianchi identity $\dd F_a+[A_a,F_a]=0$. Moreover, the curvatures of $(\{g_{ab}\},\{A_a\})$ and $(\{\tilde{g}_{ab}\},\{\tilde{A}_a\})$ are related via
\begin{equation}
 \tilde F_a\ =\ g_a^{-1} F_a g_a~~~\mbox{on}~~~U_a~.
\end{equation}

\paragraph{Degree 2.} A degree-2 cochain with values in the Lie crossed module $(\sH\stackrel{\sft}{\to}\sG)$ consists of the following \v{C}ech cochains:
\begin{equation}
\begin{aligned}
 &\{h_{abc}\}\ \in\ C^{2,0}(\frU,\sH)~,~~~&&\{\Lambda_{ab}\}\ \in\ C^{1,1}(\frU,\frh)~,~~~\{B_a\}\ \in\ C^{0,2}(\frU,\frh)~,\\
 &\{g_{ab}\}\ \in\ C^{1,0}(\frU,\sG)~,~~~&&\{A_a\}\ \in\ C^{0,1}(\frU,\frg)~.
\end{aligned}
\end{equation}
Note that contrary to Deligne cohomology, the sum of \v{C}ech and de Rham degrees of the \v{C}ech cochains forming the non-Abelian differential cochain will no longer be constant from now on. The degrees are constant, however, across the form-valued \v{C}ech cochains which take values in the same Lie algebra or integrating Lie group. Moreover, since we fixed our conventions such that $g_{aa}=\unit$, we do not have the additional elements $C^{0,0}(\frU,\sH)$ that appeared in Schreiber \& Waldorf \cite{Schreiber:2008aa}. The degree-2 cocycle conditions are
\begin{subequations}
\begin{eqnarray}
 h_{acd}h_{abc} \! &=&\!  h_{abd}(g_{ab}\acton h_{bcd})~,\\
 \Lambda_{ac}\! &=&\! \Lambda_{bc}+g_{bc}^{-1}\acton\Lambda_{ab}-g_{ac}^{-1}\acton(h_{abc}\nabla_ah_{abc}^{-1})~,\\
 B_b\! &=&\!   g^{-1}_{ab}\acton B_a -\nabla_b\Lambda_{ab}-\tfrac{1}{2}[\Lambda_{ab},\Lambda_{ab}]~,\\
g_{ac} \! &=&\!  \sft(h_{abc})g_{ab}g_{bc}~,\\
 A_b\! &=&\! g_{ab}^{-1}A_a g_{ab}+g_{ab}^{-1}\dd g_{ab}-\sft(\Lambda_{ab})~,
\end{eqnarray}
\end{subequations}
where each equation is considered on the obvious intersections of patches. Note that upon putting $h_{aba}=\unit$, the cocycle condition for $\Lambda_{ab}$ turns into the corresponding consistency condition given in \cite{Baez:2010ya}. 

Two cocycles $(\{h_{abc}\},\{\Lambda_{ab}\},\{B_a\},\{g_{ab}\},\{A_a\})$ and $(\{\tilde h_{abc}\},\{\tilde \Lambda_{ab}\},\{\tilde B_a\},\{\tilde g_{ab}\},\{\tilde A_a\})$ are considered cohomologous, if there is a degree-1 $(\sH\stackrel{\sft}{\to}\sG)$-valued cochain
\begin{equation}
 \{h_{ab}\}\ \in\ C^{1,0}(\frU,\sH)~,~~~ \{\Lambda_a\}\ \in\ C^{0,1}(\frU,\frh)\eand \{g_a\}\ \in\ C^{0,0}(\frU,\sG)
\end{equation}
such that
\begin{subequations}\label{eq:equiv-degree-2}
 \begin{eqnarray}
 \tilde h_{abc}\! &=&\! g_a^{-1}\acton(h_{ac}h_{abc}(g_{ab}\acton h^{-1}_{bc})h^{-1}_{ab})~,\\
 \tilde \Lambda_{ab}\! &=&\! g^{-1}_b\acton \Lambda_{ab}+\Lambda_b-\tilde{g}^{-1}_{ab}\acton \Lambda_a-(g_b^{-1}g_{ab}^{-1})\acton (h_{ab}^{-1}\nabla_b h_{ab})~,\\
 \tilde{B}_a\! &=&\! g_a^{-1}\acton B_a -\tilde{\nabla}_a\Lambda_a-\tfrac{1}{2}[\Lambda_a,\Lambda_a]~,\\
 \tilde{g}_{ab}\! &=&\! g_a^{-1}\sft(h_{ab})g_{ab} g_b~,\\
  \tilde{A}_a\! &=&\! g_a^{-1} A_a g_a+g_a^{-1} \dd g_a-\sft(\Lambda_a)~.
 \end{eqnarray}
\end{subequations}
On each patch $U_a$, we introduce the curvatures
\begin{equation}
 F_a\ :=\ \dd A_a+\tfrac{1}{2}[A_a,A_a]\eand H_a\ :=\ \dd B_a+A_a\acton B_a\ =\ \nabla_a B_a~.
\end{equation}
To render the underlying parallel transport of one-dimensional objects along surfaces reparameterisation invariant, one has to impose the fake curvature condition:
\begin{equation}
 \CF_a\ :=\ F_a-\sft(B_a)\ =\ 0~.
\end{equation}
Besides the Bianchi identity for $F_a$, the fake curvature condition yields the Bianchi identity $\dd H_a+A_a\acton H_a=0$ together with
\begin{equation}
 \sft(H_a)\ =\ 0~,
\end{equation}
and this equation, together with the Peiffer identity, implies that $H_a$ takes values in the centre of $\frh$. The curvatures of cohomologous 2-cochains $(\{h_{abc}\},\{\Lambda_{ab}\},\{B_a\},\{g_{ab}\},\{A_a\})$ and $(\{\tilde h_{abc}\},\{\tilde \Lambda_{ab}\},\{\tilde B_a\},\{\tilde g_{ab}\},\{\tilde A_a\})$ are related as follows:
\begin{subequations}
 \begin{eqnarray}
 \tilde{F}_a\! &=&\! g_a^{-1} F_a g_a-\sft(\tilde{\nabla}_a\Lambda_a+\tfrac12[\Lambda_a,\Lambda_a])~,\\
 \tilde{H}_a\! &=&\! g_a^{-1}\acton H_a-(\tilde F_a-\sft(\tilde B_a))\acton \Lambda_a~.
 \end{eqnarray}
\end{subequations}

Note that the degree-2 cohomology set arose from the degree-1 set by categorification: the cocycle and coboundary relations for degree 1 hold in degree 2 only up to isomorphisms. Therefore we expect that beyond the equivalence relation between cochains, there should be an equivalence relation between equivalence relations. Two degree-0 cochains $(\{g_a\},\{h_{ab}\})$ and $(\{\tilde g_a\},\{\tilde h_{ab}\})$ encoding an equivalence relation \eqref{eq:equiv-degree-2} are called cohomologous, if there is a degree-0 cochain $\{h_a\}\in C^{0,0}(\frU,\frh)$ such that
\begin{equation}\label{eq:equiv_of_equivs}
 \begin{aligned}
  \tilde g_a\ =\ \sft(h_a)g_a\eand \tilde{h}_{ab}\ =\ h_ah_{ab}(g_{ab}\acton h_b^{-1})~.
 \end{aligned}
\end{equation}

\paragraph{Degree 3.} Degree-3 cochains with values in the Lie 2-crossed module $(\sL\overset{\sft}{\to}\sH\overset{\sft}{\to}\sG)$ are encoded in the following \v{C}ech cochains:
\begin{equation}
\begin{aligned}
 &\{\ell_{abcd}\}\in C^{3,0}(\frU,\sL)~,&&\{\Xi_{abc}\}\in C^{2,1}(\frU,\frl)~,~~\{\Sigma_{ab}\}\in C^{1,2}(\frU,\frl)~,~~\{C_a\}\in C^{0,3}(\frU,\frl)~,\\
 &\{h_{abc}\}\in C^{2,0}(\frU,\sH)~,&&\{\Lambda_{ab}\}\in C^{1,1}(\frU,\frh)~,~~~\{B_a\}\in C^{0,2}(\frU,\frh)~,\\
 &\{g_{ab}\}\in C^{1,0}(\frU,\sG)~,&&\{A_a\}\in C^{0,1}(\frU,\frg)~.
\end{aligned}
\end{equation}
The degree-3 cocycle conditions for $\{\ell_{abcd}\}$, $\{h_{abc}\}$, and $\{g_{ab}\}$ are given in \eqref{eq:CoCycle3Bun}. The corresponding equations on the gauge potentials $\{C_a\}$, $\{B_a\}$, and $\{A_a\}$ are given by gauge transformations across overlaps of patches:
\begin{subequations}\label{eq:cocycle-degree-3}
 \begin{eqnarray}
C_b\! &=&\! g^{-1}_{ab}\acton C_a-\nabla_{b}\big(\Sigma_{ab}-\tfrac12\{\Lambda_{ab},\Lambda_{ab}\}\big)+\sft(\Lambda_{ab})\acton \tfrac12\{\Lambda_{ab},\Lambda_{ab}\}+ \notag\\
  &&\hspace{1cm}+\,\{B_{b},\Lambda_{ab}\}+\{\Lambda_{ab},B_{b}\}+\{\Lambda_{ab},\nabla_{b}\Lambda_{ab}+\tfrac12[\Lambda_{ab},\Lambda_{ab}]\}~,\\
 B_b\! &=&\! g^{-1}_{ab}\acton B_a-\nabla_b\Lambda_{ab}-\tfrac12[\Lambda_{ab}, \Lambda_{ab}]-\sft(\Sigma_{ab})~,\\
 A_b\! &=&\! g^{-1}_{ab}A_{a} g_{ab}+g^{-1}_{ab}\dd g_{ab}-\sft(\Lambda_{ab})~.
 \end{eqnarray}
\end{subequations}
The degree-3 cocycle condition for $\{\Lambda_{ab}\}$ is an obvious categorification of the degree-2 cocycle condition of $\{\Lambda_{ab}\}$. We omit the remaining cocycle conditions for $\{\Xi_{abc}\}$ and $\{\Sigma_{ab}\}$, as their explicit form is mostly irrelevant for working with higher gauge theories based on principal 3-bundles. Moreover, they are easily derived: the  degree-3 cocycle condition for $\{\Lambda_{ab}\}$ and $\{\Sigma_{ab}\}$ are obtained by demanding consistency of the `gluing relations' \eqref{eq:cocycle-degree-3} across triple intersections of patches $U_a\cap U_b\cap U_c$. The consistency of the thus obtained cocycle conditions across quadruple intersections of patches then yields the cocycle condition for $\{\Xi_{abc}\}$.

Two degree-3 cochains $(\{\ell_{abcd}\},\{\Xi_{abc}\},\{\Sigma_{ab}\},\{C_a\},\{h_{abc}\},\{\Lambda_{ab}\},\{B_a\},\{g_{ab}\},\{A_a\})$\linebreak and $(\{\tilde \ell_{abcd}\},\{\tilde\Xi_{abc}\},\{\tilde\Sigma_{ab}\},\{\tilde C_a\},\{\tilde h_{abc}\},\{\tilde\Lambda_{ab}\},\{\tilde B_a\},\{\tilde g_{ab}\},\{\tilde A_a\})$ are cohomologous, if relations \eqref{eq:CoBoundary3Bun}, the following equations: 
\begin{subequations}
 \begin{eqnarray}
\tilde C_a\! &=&\! g^{-1}_{a}\acton C_a-\tilde\nabla_{a}^0\big(\Sigma_{a}-\tfrac12\{\Lambda_{a},\Lambda_{a}\}\big)+\notag\\
  &&\hspace{1cm}+\,\{\tilde B_{a},\Lambda_{a}\}+\{\Lambda_{a},\tilde B_{a}\}+\{\Lambda_{a},\tilde{\nabla}_{a}\Lambda_{a}+\tfrac12[\Lambda_{a},\Lambda_{a}]\}~,\\
 \tilde B_a\! &=&\! g^{-1}_{a}\acton B_a-\tilde{\nabla}_a\Lambda_{a}-\tfrac12[\Lambda_{a},\Lambda_{a}]-\sft(\Sigma_{a})~,\\
 \tilde A_a\! &=&\! g^{-1}_{a}A_{a} g_{a}+g^{-1}_{a}\dd g_{a}-\sft(\Lambda_{a})~,
 \end{eqnarray}
\end{subequations}
and additional equations for $\{\Xi_{abc}\}$, $\{\Sigma_{ab}\}$, and $\{\Lambda_{ab}\}$, which we again suppress, are satisfied. Here, $\tilde{\nabla}_a^0:=\dd+\tilde A^0_a\acton$ with $\tilde A_a^0:=\tilde A_a^0+\sft(\Lambda_a)$.

Again, to have a well-defined underlying parallel transport along volumes, the curvatures
\begin{equation}
 F_a\ :=\ \dd A_a+\tfrac{1}{2}[A_a,A_a]~,~~H_a\:=\ \nabla_a B_a~,~~G_a\ :=\ \nabla_a C_a+\{B_a,B_a\}
\end{equation}
have to satisfy the fake curvature conditions
\begin{equation}
 \CF_a\ :=\ F_a-\sft(B_a)\ =\ 0\eand \CH_a\ :=\ H_a-\sft(C_a)\ =\ 0~.
\end{equation}
Besides the usual Bianchi identity involving $F_a$, we have
\begin{equation}
 \nabla_a H_a+\sft(\{B_a,B_a\})\ =\ 0\eand \nabla_a G_a\ =\ 0
\end{equation}
together with
\begin{equation}
 \sft(H_a)\ =\ 0\eand \sft(G_a)\ =\ 0~.
\end{equation}
Once again, this should be compared to the principal 2-bundle case. Because of the Peiffer lifting, the equation $\sft(H_a)=0$ does not imply that $H_a$ takes values in the centre of $\frh$, rather it is non-Abelian in general. However, $\sft(G_a)=0$ implies that $G_a$ lies in the centre of $\frl$ since $(\frl\overset{\sft}{\to}\frh,\acton)$, with $\acton$ being the induced $\frh$-action on $\frl$ defined in \eqref{eq:InducedActionAlgebra}, is a differential crossed module. 

The curvatures of cohomologous 3-cochains are then related via
\begin{subequations}
 \begin{eqnarray}
 \tilde F_{a}\! &=&\!  g^{-1}_aF_{a} g_a-\sft\big(\tilde \nabla_{a}\Lambda_{a}+\tfrac12\sft(\Lambda_{a})\acton \Lambda_{a}\big)~,\\
\tilde H_{a}\! &=&\! g^{-1}_a\acton H_{a}-\big(\tilde F_{a}-\sft(\tilde B_{a})\big)\acton\Lambda_{a}+\notag\\
 &&\hspace{1cm}+\, \sft\Big[-\tilde \nabla_{a}^0\big(\Sigma_{a}-\tfrac12\{\Lambda_{a},\Lambda_{a}\}\big)+\{\tilde B_{a},\Lambda_{a}\}+\{\Lambda_{a},\tilde B_{a}\}+\notag\\
  &&\kern2cm+\,\{\Lambda_{a},\tilde\nabla_{a}\Lambda_{a}+\tfrac12[\Lambda_{a},\Lambda_{a}]\}\Big]~,\\
\tilde G_{a}\! &=&\! g^{-1}_a\acton G_{a}-\big(\tilde F_{a}-\sft(\tilde B_{a})\big)\acton\big(\Sigma_{a}-\tfrac12\{\Lambda_{a},\Lambda_{a}\}\big)+\notag\\
   &&\hspace{1cm}+\,\{\Lambda_{a},\tilde H_{a}-\sft(\tilde C_{a})\}-\{\tilde H_{a}-\sft(\tilde C_{a}),\Lambda_{a}\}-\notag\\
    &&\kern1cm-\,\{\Lambda_{a},\big(\tilde F_{a}-\sft(\tilde B_{a})\big)\acton\Lambda_{a}\}~.
 \end{eqnarray}
\end{subequations}

Eventually, note that there are again equivalence relations between equivalence relations here, and one has a categorified version of equation \eqref{eq:equiv_of_equivs}. They appear in the Penrose--Ward transform in \eqref{eq:equiv_of_equivs1}, but they turn out to be irrelevant for the resulting field equations, as one would expect. Note that the $h_a$s used in the categorification of \eqref{eq:equiv_of_equivs} and the ones appearing in \eqref{eq:equiv_of_equivs1} are related by $h_a \leftrightarrow g_a^{-1}\acton h_a$. In addition, there is a further equivalence relation between these equivalences. Since these formul{\ae} are rather lengthy, not very illuminating and of no direct use in the discussion of the dynamics of connective structures on principal 3-bundles, we again refrain from listing them here.

\section{Conclusions}\label{sec:Conclusions}

In this paper, we constructed new $\CN=(n,0)$ superconformal field theories in six dimensions with $n=0,1,2$ that contain a non-Abelian generalisation of the tensor multiplet. The equations were obtained from a Penrose--Ward transform of certain holomorphic principal 3-bundles over a suitable twistor space. Compared to the superconformal field equations that we had derived previously in \cite{Saemann:2012uq}, the equations here are significantly more general: the Peiffer identity is lifted in a controlled way and the previous restriction of the 3-form curvature $H$ to live in the centre of a Lie algebra is removed. Moreover, our new equations contain a 3-form potential, which can be motivated by either making connections to M-theory or by referring to other approaches to six-dimensional superconformal field theories as those in \cite{Samtleben:2011fj,Samtleben:2012mi,Samtleben:2012fb,Bandos:2013jva} or \cite{Fiorenza:2012tb}. 

The Penrose--Ward transform exposed all features of higher gauge theory with principal 3-bundles, some of which had remained  unexplored in the literature so far. In particular, we formulated the non-Abelian differential cohomology that describes principal 3-bundles with connective structure. This cohomology nicely reduces to the usual Deligne cohomology, when the principal 3-bundle is reduced to an Abelian 2-gerbe.

The constraint equations we obtained seem rather promising to us, and they lead to a number of questions that we intend to address in future work. First of all, it is important to reduce our superfield constraint equations to actual field equations on ordinary six-dimensional space-time. This issue appears usually in the twistor description of field equations, see \cite{Harnad:1984vk,Harnad:1985bc} for the case of maximally supersymmetric Yang--Mills theory in four dimensions and \cite{Samtleben:2009ts,Samtleben:2010eu} for similar expansions in the context of three-dimensional supersymmetric Chern--Simons theories. Once this is done, a more detailed analysis of the field equations and their possible relation to the effective description of M5-branes can be undertaken. In particular, the reduction to five-dimensional super Yang--Mills theory as well as a detailed study of the BPS configurations known as self-dual strings should be performed. As we pointed out, our superconformal tensor field equations can be dimensionally reduced to those of a non-Abelian generalisation of the self-dual string equations. Alternatively, these equations can also be obtained from holomorphic principal 3-bundles over the hyperplane twistor space introduced in \cite{Saemann:2011nb}.

An important question is the interpretation of the additional 3-form potential that is not believed to be part of the field content of an effective description of M5-branes. Currently, it seems that this field should be regarded analogously to the 3-form field appearing in \cite{Samtleben:2011fj,Samtleben:2012mi,Samtleben:2012fb,Bandos:2013jva}. That is, it merely mediates couplings between the various other fields.

Finally, it still seems conceivable that a manageable \v{C}ech description of principal 2-bundles with semistrict structure 2-groups exists. In this case, our twistor construction would be an ideal approach both to explore the general definition of semistrict higher gauge theories as well as to find new and more general superconformal field theories in six dimensions.

\section*{Acknowledgements}
We would like to thank D.~Baraglia, D.~Berman, C.~S.~Chu, P.~Hekmati, B.~Jur{\v c}o, N.~Lambert, L.~Mason, A.~Mikovi\'c, M.~Murray, G.~Papadopoulos, D.~Roberts, D.~Smith, D.~Sorokin, and R.~Vozzo for enjoyable discussions. CS was supported by an EPSRC Career Acceleration Fellowship.

\appendices

\subsection{Collection of Lie 2-crossed module identities and their proofs}\label{sec:Proofs}

\paragraph{Useful Lie 2-crossed module identities.}
Let us collect and prove some useful identities which are used throughout this work.

We have 
\begin{equation}
 \{\unit,h\}\ =\ \unit\ =\ \{h,\unit\}\eforall h\ \in\ \sH~,
\end{equation}
as follows directly from applying the Lie 2-crossed module axioms (iv) and (v) to $\{\unit\unit,h\}$ and $\{h,\unit\unit\}$, respectively.

Obvious, but very useful is also
\begin{equation}
 \ell_1\ell_2\ =\ \{\sft(\ell_1),\sft(\ell_2)\}\ell_2\ell_1\eforall \ell_1,\ell_2\ \in\ \sL~.
\end{equation}

By applying the Lie 2-crossed module axiom (iv) to the expression $\{h_1h_1^{-1},h_1h_2h_1\}$, we find
\begin{equation}
  \{h_1,h_2\}^{-1}\ =\ \sft(h_1)\acton\{h_1^{-1},h_1h_2h_1^{-1}\}\eforall h_1,h_2\ \in\ \sH~.
\end{equation}
Together, with axiom (vi), this enables us to rewrite the induced $\sH$-action on $\sL$ given in  \eqref{eq:ActHToL} as
\begin{equation}\label{eq:alt_def_H_on_L}
 h\acton\ell\ =\ \sft(h)\acton(\ell\{h^{-1},h\sft(\ell^{-1})h^{-1}\})\eforall h\ \in\ \sH\eand\ell\ \in\ \sL~.
\end{equation}

Note that
\begin{eqnarray}\label{eq:commPB}
 \{h_1,h_2\}\ell\! &=&\! \ell\ell^{-1}\{h_1,h_2\}\ell(\{h_1,h_2\})^{-1}\{h_1,h_2\}\notag \\
 \! &=&\!  \ell\{\sft(\ell)^{-1},\sft(\{h_1,h_2\})\}\{h_1,h_2\}\notag\\
 \! &=&\!  (\sft(\{h_1,h_2\})\acton \ell)\{h_1,h_2\}
\end{eqnarray}
for all $h_1,h_2\in\sH$ and $\ell\in\sL$.

As observed in \cite{Martins:2009aa}, the fact that the action of $\sH$ onto $\sL$ is an automorphism together with \eqref{eq:commPB} implies that we can reformulate the Lie 2-crossed module axiom (v) as follows:
\begin{eqnarray}\label{eq:formulaAxiom5}
  \{h_1,h_2h_3\}\! &=&\! \{h_1,h_2\}\{h_1,h_3\}\{\langle h_1,h_3\rangle^{-1},\sft(h_1)\acton h_2\}\notag\\
  \! &=&\!  \{h_1,h_2\}((\sft(h_1)\acton h_2)\acton\{h_1,h_3\})\notag\\
  \! &=&\!  (\sft(\{h_1,h_2\})\acton(\sft(h_1)\acton h_2)\acton\{h_1,h_3\})\{h_1,h_2\}\notag\\
  \! &=&\!  ((h_1h_2h_1^{-1}(h_1h_2^{-1}h_1^{-1}))\acton(h_1h_2h_1^{-1})\acton\{h_1,h_3\})\{h_1,h_2\}\notag\\
 \! &=&\! ((h_1h_2h_1^{-1})\acton\{h_1,h_3\})\{h_1,h_2\}
\end{eqnarray}
for all $h_1,h_2,h_3\in\sH$.

\paragraph{Proof that the induced $\sH$-action is by automorphisms.}
To verify that the induced $\sH$-action  \eqref{eq:ActHToL} is an automorphism action, one has to demonstrate the two relations $h_1h_2\acton \ell=h_1\acton h_2\acton \ell$ and $h\acton (\ell_1\ell_2)=(h\acton \ell_1)(h\acton \ell_2)$ for all $h,h_1,h_2\in\sH$ and $\ell,\ell_1,\ell_2\in\sL$.

To show the first relation, we use the alternative version \eqref{eq:alt_def_H_on_L} for the induced $\sH$-action on $\sL$. We find
\begin{eqnarray}
  (h_1h_2)\acton\ell\ \! &=&\! (\sft(h_1h_2)\acton\ell)\{\sft(h_1h_2)\acton (h_1h_2)^{-1},\sft(h_1h_2)\acton(h_1h_2\sft(\ell^{-1})h_2^{-1}h_1^{-1})\}\notag\\
  \! &=&\! (\sft(h_1h_2)\acton\ell)\{\sft(h_1h_2)\acton h_2^{-1},\sft(h_1h_2)\acton(h_2\sft(\ell^{-1})h_2^{-1})\}\times\notag\\
  &&\hspace{2cm}\times\,\{\sft(h_1)\acton h_1^{-1},\sft(h_1)\acton(h_1h_2\sft(\ell^{-1})h_2^{-1}h_1^{-1}\}\notag \\
  \! &=&\! h_1\acton(h_2\acton \ell)~.
\end{eqnarray}
To show the second relation, we consider
\begin{eqnarray}
  (h\acton \ell_1)(h\acton \ell_2)\! &=&\!\ell_1\{\sft(\ell_1^{-1}),h\}\ell_2\{\sft(\ell_2^{-1}),h\}\notag\\
  \! &=&\! \ell_1\ell_2\{\sft(\ell_2^{-1}),\sft(\ell_1^{-1})h\sft(\ell_1)h^{-1}\}\{\sft(\ell_1^{-1}),h\}\{\sft(\ell_2^{-1}),h\}\notag\\
  \! &=&\! \ell_1\ell_2\{\sft(\ell_2^{-1}),\sft(\ell_1^{-1})h\sft(\ell_1)h^{-1}\}\times\notag\\
  &&\hspace{1cm}\times\,\{\sft(\ell_2^{-1}),h\}\{h\sft(\ell_2^{-1})h^{-1}\sft(\ell_2),\sft(\ell_1^{-1})h\sft(\ell_1)h^{-1}\}\{\sft(\ell_1^{-1}),h\}\notag\\
  \! &=&\! \ell_1\ell_2\{\sft(\ell_2^{-1}),\sft(\ell_1^{-1})h\sft(\ell_1)h^{-1}h\}\{\sft(\ell_1^{-1}),h\}\notag\\
  \! &=&\! \ell_1\ell_2\{\sft(\ell_2^{-1}),\sft(\ell_1^{-1})h\sft(\ell_1)\}\{\sft(\ell_1^{-1}),h\}\notag\\
  \! &=&\! \ell_1\ell_2\{\sft(\ell_2^{-1})\sft(\ell_1^{-1}),h\}\notag\\
  \! &=&\! h\acton (\ell_1\ell_2)~.
\end{eqnarray}

\pagebreak



\end{document}